\documentclass[12pt]{article}
\usepackage{amssymb,amsmath}
\usepackage{latexsym}
\usepackage{graphicx}
\usepackage{color}
\usepackage{datetime}
\usepackage{cite}
\usepackage[
      colorlinks=false,
      linkcolor=darkblue,  
      urlcolor=blue,    
      filecolor=blue,     
      citecolor=red,
      linktocpage=true,
      pdfstartview=FitV,
      bookmarksopen=true    
      ]{hyperref}

\usepackage{psfrag}

\numberwithin{equation}{section} 

\makeatletter
\g@addto@macro\bfseries{\boldmath}
\makeatother

\definecolor{cardinal}{rgb}{0.6,0,0}
\definecolor{darkgreen}{rgb}{0,0.5,0}
\definecolor{golden}{rgb}{0.92, 0.7, 0}
\definecolor{midnight}{rgb}{0, 0, 0.5}
\definecolor{darkblue}{rgb}{0.2, 0, 0.8}

\def\Neql#1{{\cal N}\!=\!{#1}}
\def\coeff#1#2{\relax{\textstyle {#1 \over #2}}\displaystyle}

\def\ZZ{\mathbb{Z}}

\def\cB{{\cal B}}

\def\cD{{\cal D}}
\def\cF{{\cal F}}
\def\cG{{\cal G}}

\def\cL{{\cal L}}

\def\cN{{\cal N}}
\def\cP{{\cal P}}

\def\cR{{\cal R}}
\def\cS{{\cal S}}

\def\nBPS#1{$\frac{1}{#1}$-BPS}

\def\cO{{\cal O}}
\def\jt{\tilde{j}}
\def\Jt{\tilde{J}}

\def\psit{\widetilde{\psi}}

\def\bbC{\mathbb{C}}
\def\bbR{\mathbb{R}}
\def\bbZ{\mathbb{Z}}

\def\cLh{\widehat{\mathcal{L}}}

\def\ket#1{|{#1}\rangle}
\def\half{\frac{1}{2}}

\begin{document}

\begin{titlepage}

\vspace*{-11ex}
\begin{flushright}
DFPD-15-TH-06\\
IPhT-T15/019\\
QMUL-PH-15-05\\
YITP-15-12
\end{flushright}

\bigskip
\centerline{\LARGE \bf Habemus Superstratum!}
\medskip
\centerline{\Large \bf  A constructive proof of the existence of superstrata}

\bigskip\bigskip
\centerline{{\bf Iosif Bena,$^1$ Stefano Giusto,$^{2,3}$ Rodolfo Russo,$^4$
Masaki Shigemori$^{5,6}$ and Nicholas P. Warner$^{7}$}}
\bigskip
\centerline{$^1$ Institut de Physique Th\'eorique, }
\centerline{CEA Saclay, F-91191 Gif sur Yvette, France}
\medskip
\centerline{$^2$ Dipartimento di Fisica ed Astronomia ``Galileo Galilei''}
\centerline{Universit\`a di Padova, Via Marzolo 8, 35131 Padova, Italy}
\medskip
\centerline{$^3$ I.N.F.N. Sezione di Padova, Via Marzolo 8, 35131 Padova, Italy}
\smallskip
\centerline{$^4$ Centre for Research in String Theory, School of Physics and Astronomy}
\centerline{Queen Mary University of London, Mile End Road, London, E1 4NS,
United Kingdom}
\medskip
\centerline{$^5$ Yukawa Institute for Theoretical Physics, Kyoto University}
\centerline{Kitashirakawa-Oiwakecho, Sakyo-ku, Kyoto 606-8502 Japan}
\medskip
\centerline{$^6$ Hakubi Center, Kyoto University}
\centerline{Yoshida-Ushinomiya-cho, Sakyo-ku, Kyoto 606-8501, Japan}
\medskip
\centerline{$^7$ Department of Physics and Astronomy,}
\centerline{University of Southern California,} \centerline{Los
Angeles, CA 90089, USA}
\bigskip

\begin{abstract}
\noindent
We construct the first example of a superstratum: a class of
smooth horizonless supergravity solutions that are parameterized
by arbitrary continuous functions of (at least) two variables and
have the same charges as the supersymmetric D1-D5-P black
hole.   We work in Type IIB string theory on $T^4$ or $K3$ and our
solutions involve a subset of fields that can be described by a
six-dimensional supergravity with two tensor multiplets. The
solutions  can thus be constructed using a linear structure, and 
we give an explicit recipe to start from a superposition of modes 
specified by an arbitrary function of two
variables and impose regularity to obtain the full horizonless
solutions in closed form.  We also give the precise CFT
description of these solutions and show that they are not dual
to descendants of chiral primaries. They are thus 
much more general than all the known solutions whose CFT dual is
precisely understood. Hence our construction represents a
substantial step toward the ultimate goal of constructing the
fully generic superstratum that can account for a finite fraction
of the entropy of the three-charge black hole in the regime of parameters 
where the classical black hole solution exists.
\end{abstract}

\end{titlepage}


\tableofcontents

\newpage
\section{Introduction}
\label{Sect:introduction}

There has been growing evidence that string theory contains smooth,
horizonless bound-state or solitonic objects that have the same charges
and supersymmetries as large BPS black holes and that depend on
arbitrary continuous functions of two variables. These objects, dubbed
\emph{superstrata}, were first conjectured to exist in
\cite{Bena:2011uw}, by realizing that some of the exotic brane bound
states studied in \cite{deBoer:2010ud}\footnote{In
\cite{deBoer:2010ud}, double supertube transitions \cite{Mateos:2001qs}
of branes were argued to lead to configurations that are parametrized by
functions of two variables and are generically non-geometric.  For
further developments on exotic branes see \cite{deBoer:2012ma}.} can
give rise to non-singular solutions in the duality frame where the
charges of these objects correspond to momentum, D1-branes and
D5-branes.

It was subsequently argued that, assuming that
superstrata existed, the most general class of such objects could carry 
an entropy that scales with the charges in exactly the same way as the
entropy of the D1-D5-P black hole, and possibly even with the same
coefficient \cite{Bena:2014qxa}. Since this entropy would come entirely from smooth
horizonless solutions, this would substantiate the 
fuzzball description of supersymmetric black holes in string theory:
the classical solution describing these black holes stops giving a
correct description of the physics at the scale of the horizon, where
a new description in terms of fluctuating superstrata geometries takes
over.

Partial evidence for the existence of superstrata can be obtained by
analyzing string emission in the D1-D5 system \cite{Giusto:2011fy,Giusto:2012jx}, or by constructing certain smaller classes of
supergravity solutions \cite{Giusto:2013bda,Niehoff:2012wu,Lunin:2012gp,Vasilakis:2013tjs,Niehoff:2013kia,Shigemori:2013lta}. However, to prove that superstrata indeed exist, one needs to explicitly construct smooth horizonless solutions that have the same charges as the D1-D5-P black hole and are parameterized by arbitrary continuous functions of two variables, which is a challenging problem.
  
The purpose of this paper is to construct such solutions and thus
demonstrate that superstrata exist.  Furthermore, we will be able to
find precisely the CFT states dual to these solutions and  show that
these states are not descendants of chiral primaries, which means that
they are much more general than all the known solutions whose CFT dual
is precisely understood \cite{Giusto:2004id,
Ford:2006yb, Lunin:2012gp, Giusto:2013bda}. This is a huge step toward achieving the
ultimate goal of constructing all smooth horizonless solutions that have
the right properties for reproducing the black-hole entropy and thus
proving the fuzzball conjecture for BPS black holes.

Our procedure relies on the proposal \cite{Bena:2011uw} that superstrata
can be obtained by adding momentum modes on two-charge D1-D5 supertubes:
Supertube solutions \cite{Lunin:2001fv, Emparan:2001ux, Lunin:2002iz}
have eight supercharges and are parameterized by functions of one
variable; adding another arbitrary function-worth of momentum modes to
each supertube was argued to break the supersymmetry to four
supercharges and result in a superstratum parameterized by arbitrary
continuous functions of \emph{two} variables. However, as anybody
familiar with supertube solutions might easily guess, trying to follow
this route brings one rather quickly into a technical quagmire.

A simpler route to prove that superstrata exist is to start from a
maximally-rotating supertube solution and try to deform this solution by
making the underlying fields and metric wiggle in two directions. This
approach is attractive for several reasons.  First, the holographic
dictionary for the \nBPS{4} (8-supercharge) \footnote{Throughout this
paper \nBPS{N} will denote a state with $\frac{32}{N}$ supercharges.}
D1-D5 supertubes is well understood
\cite{Kanitscheider:2006zf,Kanitscheider:2007wq} and so, as we will
describe later in this paper, we can then generalize this dictionary to
the \nBPS{8} (4-supercharge) D1-D5-P superstrata. Second, the equations
that govern the superstrata solutions are well-known
\cite{Gutowski:2003rg,Cariglia:2004kk}, and can be organized in a linear
fashion \cite{Bena:2011dd}, and so this technique appears to be the
technique of choice, all the more so because it has enabled the
construction of solutions that depend of two arbitrary functions each of
which depends upon a different variable
\cite{Niehoff:2013kia}. Nevertheless, while extensive trial and error
has led to many solutions that depend on functions of two variables they
have all, so far, been singular\footnote{It is important to remember
that our purpose is to reproduce the black hole entropy by counting
\emph{smooth} horizonless supergravity solutions, or at most singular
limits thereof, that one can honestly claim to describe in a
controllable way. If we were to count black hole microstate solutions
with singularities, we could easily overcount the entropy of many a
black hole.}.

The key ingredient simplifying the task of smoothing the singularities of these solutions
is a fourth type of electric field that appears neither in the original
five-dimensional $U(1)^3$ ungauged supergravity, where most of the known
black hole microstate solutions have been built \cite{Bena:2005va,
Berglund:2005vb, Bena:2006kb}, nor in the six-dimensional uplift in
\cite{Gutowski:2003rg, Bena:2011dd}, where the solutions of
\cite{Niehoff:2013kia} were constructed.  The presence of this field can
drastically simplify the sources that appear on the right-hand sides of
the equations governing the superstratum and allows us to find smooth
solutions depending on functions of two variables in closed form. The
solutions with this field can only be embedded in a five-dimensional
ungauged supergravity with four or more $U(1)$ factors, or in a
six-dimensional supergravity with two or more tensor
multiplets. Fortunately, the equations underlying the most general
supersymmetric solution of the latter theory were found in
\cite{Giusto:2013rxa} and these equations can also be solved following a
linear algorithm similar to the one found in \cite{Bena:2011dd}.

The essential role for this fourth type of electric field in the
solutions dual to the typical microstates of the D1-D5-P black hole was
first revealed by analyzing string emission from the D1-D5-P system
\cite{Giusto:2011fy,Giusto:2012jx} and from D1-D5 precision holography
\cite{Kanitscheider:2006zf,Kanitscheider:2007wq}. Furthermore, in
\cite{Bena:2013ora,Bena:2014rea} it was shown that adding this field to
certain fluctuating supergravity solutions can make their singularities
much milder\footnote{This has allowed, for example, the construction of
an infinite-dimensional family of black ring solutions that gives the
largest known violation of black-hole uniqueness in any theory with
gravity \cite{Bena:2014rea}.}.  The fact that the extra field plays an
important part in both obtaining smooth, fluctuating three-charge
geometries and in the description of D1-D5-P string emission processes is,
in our opinion, no coincidence, but rather an indication that the 
solutions we construct are necessary ingredients in the description of the typical
microstates of the three-charge black hole.

Our plan is  to start from a round supertube solution with the
fourth electric field turned on and to prove that this solution is part
of a family of solutions that is parameterized by functions of two
variables. There are two natural perspectives on these solutions.

The first is to recall that, in the D1-D5 duality frame, the infra-red geometry of the two-charge supertube solution is AdS$^{\rm global}_3 \times S^3$. This background has three $U(1)$ symmetries, which we will parametrize by $(v,\psi,\phi)$: $v$ corresponds to the D1-D5 common direction, $\psi_{GH}\equiv \psi+\phi$
 is the Gibbons-Hawking fiber that comes from writing the $\mathbb{R}^4$ in which supertube lives as a Gibbons-Hawking space and $\phi_{GH}\equiv \psi-\phi$ is the angular coordinate in the Gibbons-Hawking $\mathbb{R}^3$ base. The Lunin-Mathur two-charge supertube solutions \cite{Lunin:2001fv, Lunin:2002iz}, as well as their generalizations that have the fourth type of electric field turned on \cite{Kanitscheider:2006zf,Kanitscheider:2007wq,Giusto:2013bda}, correspond to shape deformations of the supertube, and their shapes and charge densities can be viewed as being determined  by arbitrary functions of the coordinate $\psi_{GH}$.   One can also construct solutions that depend on $v$ by simply interchanging $v$ and $\psi_{GH}$ \cite{Niehoff:2013kia}. Both these classes of solutions are parameterized by functions of one variable and, as such, correspond to special choices of spherical harmonics on the three-sphere of the round supertube solution. Our superstrata will depend non-trivially upon all three angular coordinates, but only through a two-dimensional lattice of mode numbers (defined in (\ref{modelattice})).

The second perspective comes from  decomposing the functions of two variables that parametrize our superstratum solutions under the $SU(2)_L\times SU(2)_R$ isometry of the $S^3$ and the $SL(2,\mathbb{R})_L \times
SL(2,\mathbb{R})_R$ isometry of the AdS$_3$. The shape
modes of the two-charge supertube preserve eight supercharges and have $SU(2)_L\times SU(2)_R$ quantum numbers
$(J^3,\Jt^3)=(j,\jt)$ and $SL(2,\bbR)_L\times SL(2,\bbR)_R$ weights\footnote{Here we are considering the
Ramond-Ramond (RR) sector.} $h=\tilde{h}=0$; since $|j-\jt|$ determines the spin of
the field in the theory, each Fourier mode is determined essentially by
one quantum number.  Thus, these solutions are parameterized by
functions of one variable, as expected.

The solutions we construct have four supercharges and correspond to
adding left-moving momentum modes to the supertube. The generic mode
will have $SL(2,\bbR)_L$ weight $h>0$. Since $h$ is independent of $j$,
these will generate intrinsically two-dimensional shape modes on the
$S^3$.  Since the equations underlying our solutions can be solved using
a linear algorithm, superposing multiple spherical harmonics gives rise
to very complicated source terms in the equations we are trying to
solve. Furthermore, most of the solutions one finds by brute force give
rise to singularities.  In the earlier construction of microstate
geometries, such singularities were canceled by adding homogeneous
solutions to the equations. Here we will see that this technique does
not allow us to obtain smooth solutions from a generic superposition of
harmonics on the $S^3$ in all electric fields, and that we have to
relate the combinations of spherical harmonics appearing in the electric
fields.  At the end of the day, the resulting smooth solutions will
contain one general combination of spherical harmonics on a
three-sphere, which can be repackaged into an arbitrary continuous
function of two variables.

Our superstratum can be precisely identified with a state at the free
orbifold point of the D1-D5 CFT\@. The dual CFT interpretation, besides
providing a crucial guide for the supergravity construction, firmly
establishes that our solutions contribute to the entropy of the
three-charge black hole, and clarifies what subset of the microstate
ensemble is captured by our solutions. In the previous literature, all
three-charge geometries with a known CFT dual \cite{Giusto:2004id,
Ford:2006yb, Lunin:2012gp, Giusto:2013bda} had been obtained by acting on a
two-charge solution (in the decoupling limit) with a coordinate
transformation that does not vanish at the AdS$_3$ boundary. On the CFT
side this is equivalent to acting with an element of the chiral algebra
on a Ramond-Ramond (RR) ground state, and produces a state which is
identified with a descendant of a chiral primary state in the
Neveu-Schwarz--Neveu-Schwarz (NSNS) sector.  In contrast, the
microstate solutions we construct here cannot be related, generically, to
two-charge microstate solutions via a global chiral algebra rotation. They thus
do not correspond to descendants of chiral primaries but represent much
more generic states than the ones previously considered in
\cite{Giusto:2004id,Ford:2006yb,Lunin:2012gp,Giusto:2013bda}.

In the interests of full disclosure, while the results presented here represent a major step forward in the microstate geometry programme, it is also very important to indicate what we have not yet achieved.  

First, the superstratum solutions we construct in this paper are still
rather ``coarsely grained'' in that they do not fully capture states in
the twisted sector of the dual CFT (see Section \ref{Sect:D1D5CFT}).
That is, while we do indeed have a superstratum that fluctuates
non-trivially as a function of two variables, the fluctuations we
construct here are dual to restricted classes of integer-moded
current-algebra excitations in the dual CFT and so, at present, our
superstrata solutions do not have sufficiently many states to capture
the black-hole entropy.  Thus, we have not yet achieved the ``holy
grail'' of the microstate geometry programme. 

One should also note that typical states will contain general combinations of
fractional-moded excitations in a twisted sector of very high twisting,
corresponding to a long effective string of length equal to the product
of the numbers of D1 and D5 branes.  This sector of the CFT might not be
well described within supergravity. However, to prove the validity of
the microstate geometry programme it is sufficient to show the existence
of a superstratum which contains general fractional modes in twisted
sectors of arbitrary finite order; this will establish the existence of a
mechanism which allows to encode the information of generic states in
the geometry. The fact that, in the limit of very large twisting,
corrections beyond supergravity might have to be taken into account does not invalidate the existence of such a mechanism. In particular, we hope
that in subsequent work we will be able to refine the mode analysis and
the holographic dictionary obtained in this paper and obtain superstrata
containing general fractional modes.

The other, more technical issue is that the systematic procedure given in this paper does not yet provide a complete description of the solution for all  combinations of Fourier modes of the arbitrary function of two variables that parametrizes the superstratum.  As yet, we have not been able to obtain the closed expression for one function that appears in some components of the angular momentum vector. In principle these could be singular, but we do not expect this, for two reasons:  First, we have the general explicit solution for one of the  components of the angular momentum vector and this component is regular and, from our experience, if there are singularities in the angular momentum vector they always appear in this particular component. Secondly, we have actually been able to find this function and construct the complete solution for several (infinite) families of collections of Fourier modes.  These families were chosen so as to expose possible singular behaviors and none were found.    Thus, while we do not have explicit formulae for one function that appears in the angular momentum vector for all  combinations of Fourier modes we believe that this is merely a technical limitation rather than a physical impediment.

The construction presented in this paper establishes that
the superstratum exists as a bound state object of string theory, and
that its supergravity back-reaction gives rise to smooth horizonless
three-charge solutions. Having shown this, we believe that a fully
generic superstratum is within reach and thus one will be
able to show that a finite fraction of the entropy of the BPS black hole
comes from smooth horizonless solutions. This, in turn, would imply that
the typical states of this black hole will always have a finite
component extended along the direction of the Hilbert space
parameterized by horizonless solutions, and hence will not have a
horizon. Thus one would confirm the expectations and goals of the
fuzzball/firewall arguments\footnote{See \cite{Mathur:2009hf, Bena:2007kg, Skenderis:2008qn, Balasubramanian:2008da, Chowdhury:2010ct, Almheiri:2012rt, Braunstein:2009my,Mathur:2012jk,Susskind:2012rm,Bena:2012zi,Susskind:2012uw,Avery:2012tf,Almheiri:2013hfa,Verlinde:2013uja,Maldacena:2013xja,Mathur:2013gua} for some developments in that area.}: the
horizon of an extremal supersymmetric black hole is not an essential,
fundamental component but the result of coarse-graining multiple
horizonless configurations.

More broadly, we would like to emphasize that results presented here
provide a remarkable confirmation of the power of the approach we have
been using to establish that there is structure that replaces the
horizon of a black hole: we have directly constructed this structure in
supergravity.  As we emphasized in \cite{Bena:2013dka}, this approach
could have failed at many different stages throughout its development.
The most recent hurdle has been to show that supergravity has structures
that might contain enough states to count the entropy of the black
hole. In \cite{Bena:2014qxa} we have argued that this can happen if
string theory contains three-charge superstrata solutions that can be
parameterized by arbitrary continuous functions of two variables. The
present paper shows explicitly that these solutions exist and
furthermore that they are smooth in the duality frame where the black
hole has D1,D5 and momentum charges.  (It was the successful clearing of
this latest hurdle that led to our somewhat celebratory title for this
paper.) Though most of the recent literature on the information paradox
has focused on ``Alice-and-Bob'' Gedankenexperiments, we believe that
general quantum information arguments about physics at a black-hole
horizon will always fall short of resolving the paradox: failure is
inevitable without a mechanism to support structure at the horizon
scale.  It is remarkable that string theory can provide a natural and
beautiful solution to this essential issue and, as was shown in
\cite{Gibbons:2013tqa}, microstate geometries provide the \emph{only}
possible gravitational mechanism and so must be an essential part of the
solution to the paradox.

In Section~\ref{Sect:supergravity} we introduce the six-dimensional supergravity theory where our D1-D5-P microstate solutions are constructed and also recall the connection of these solutions to those constructed in the more familiar M2-M2-M2 duality frame. We write the equations governing the supersymmetric solutions of the six-dimensional supergravity theory in a form that highlights their linear structure and simplify the problem by choosing a flat four-dimensional base space metric. The equations governing the supersymmetric solutions can then be organized in a first layer of linear equations, which determine the electric and magnetic parts of the gauge fields associated with the D1- and D5-branes, and a second layer of linear but inhomogeneous equations, which determine the momentum and the angular momentum vectors.

In Section \ref{Sect:layer1} we solve the first layer of equations.  We start from a round D1-D5 supertube carrying density fluctuations of the fourth type of electric field and apply a CFT symmetry transformation to generate a two-parameter family of modes that carry the third (momentum) charge. We then use the linearity of the equations to build solutions that contain arbitrary linear combinations of such modes. 
Section \ref{Sect:layer2} contains the most challenging technical part of the superstratum construction: finding the solution of the second layer of equations. We explain how the sources appearing in these equations have to be fine tuned to avoid singularities of the metric, and how this requirement selects a restricted set of solutions to the first layer of equations. These solutions are parameterized by certain coefficients  that can be interpreted as the Fourier coefficients of a function of two variables, which defines the superstratum. We then construct the general solution for the particular component of the angular momentum 1-form that, from our experience, controls the existence of closed timelike curves. We also find in Section \ref{Sect:examples}  the remaining components of this 1-form, thus deriving the complete solution for several (infinite) families of collections of Fourier modes. We verify  the regularity of the solutions in these examples.

Although we mostly work in the ``decoupling'' regime, in which geometries are asymptotic to AdS$_3\times S^3$, in Section \ref{Sect:physics} we present a way to extend our solutions and obtain asymptotically five-dimensional ($\,\mathbb{R}^{4,1}\times S^1$) superstrata geometries. We also derive the asymptotic charges and angular momenta of these geometries. These results are then used in Section \ref{Sect:D1D5CFT} to motivate the identification of the states dual to the superstrata at the free orbifold point of the D1-D5 CFT\@. We point out that states dual to our superstrata are descendants of non-chiral primaries and we show how some of the features of the gravity solution have a natural explanation in the dual CFT\@. 

Section \ref{Sect:Concl} summarizes the relevance of our construction for the black-hole microstate geometry programme and highlights possible future developments. Several technical results are collected in the Appendices. In Appendix \ref{Appendix:2charge} we recall the form of general two-charge microstates and in Appendix \ref{Appendix:recursion} we explain how to use a recursion relation to solve some of the differential equations of the second layer. 

Readers who are not so interested in the gory technical details of our solutions can simply read Sections ~\ref{Sect:supergravity}  and  \ref{Sect:layer1}  in order to understand the supergravity structure that we use in constructing the explicit superstratum solution, and read Section \ref{Sect:D1D5CFT} in order to understand the corresponding states in the dual CFT\@.

\section{Supergravity background}
\label{Sect:supergravity}

The existence of the superstratum was originally conjectured based upon
an analysis of supersymmetric bound states within string theory.  The (\nBPS{2}) exotic branes of string theory were thoroughly analyzed in
\cite{deBoer:2010ud, deBoer:2012ma}, where it was also argued that objects carrying dipole charges corresponding to such branes can result from simple or double supertube transitions. In \cite{Bena:2011uw} it was pointed out that the hallmark of these bound state objects is that they are locally \nBPS{2}, but when they bend to form a supertube they break some of the supersymmetry. In particular the objects that result from a simple supertube transition are \nBPS{4} and are parameterized by arbitrary functions of one variable, while the objects that result from a double supertube transition are \nBPS{8} and are parameterized by arbitrary functions of two variables.
As explained in \cite{deBoer:2010ud, deBoer:2012ma}, most of the double supertube transitions result in objects carrying exotic brane charges, which are therefore non-geometric. However, in \cite{Bena:2011uw} it was pointed out that when D1 branes, D5 branes and momentum undergo a double supertube transitions the resulting \nBPS{8} object is not only geometric but also potentially giving rise to a class of smooth microstate geometries parameterized by arbitrary functions of two variables. This object became known as the \emph{superstratum}. 
Thus, this fundamental bound state in
string theory could, as a microstate geometry, provide a very large
semi-classical contribution to the \nBPS{8} black-hole entropy.  Indeed
it was argued in \cite{Bena:2014qxa} that a fully generic superstratum
could capture the entropy to at least the same parametric growth with
charges as that of the three-charge black hole.  Thus the construction
of a completely generic superstratum has become a central goal of the
microstate geometry programme.

The supertube transitions  that yield the superstratum were analyzed in detail in  \cite{Bena:2011uw} and it was shown that indeed such solitons could be given shape modes as a function of  \emph{two} variables while remaining \nBPS{8}.  Based on the forms of these supertube transitions it was argued that the resulting geometry should be smooth but this remained to be  substantiated through computation of the fully-back-reacted geometries in supergravity.  Since this initial conjecture, much progress has been made in finding the supergravity description of the superstratum. 
 
The structure of the BPS equations led to the construction of doubly
fluctuating, but singular BPS, ``superthreads and supersheets'' in
\cite{Niehoff:2012wu,Vasilakis:2013tjs}.  Simple but very restricted classes of
superstrata were obtained in \cite{Niehoff:2013kia}.  In parallel with
this, string amplitudes were used to very considerable effect to find
the key perturbative components of the superstratum \cite{Giusto:2011fy,
Giusto:2012gt, Giusto:2012jx, Shigemori:2013lta, Giusto:2013bda}.  The
fact that the BPS equations underlying the superstratum are largely
linear \cite{Bena:2011dd} means that knowledge of the perturbative
pieces can be sufficient for generating the complete solution. Finally,
in an apparently unrelated investigation of new classes of microstate
geometries \cite{Bena:2013ora} and new families of black-ring solutions
\cite{Bena:2014rea}, a mechanism arising out of the perturbative
superstrata programme was used to resolve singularities and find new
physical solutions.

We are now in a position to pull all these threads together and obtain, for the first time, a non-trivial, fully-back-reacted smooth supergravity superstratum that fluctuates as a function of two variables.    We begin by reviewing the basic supergravity equations that need to be solve, starting in the D1-D5-P duality frame and discussing how this reduces to an analysis within six-dimensional supergravity. While we will be working with the $T^4$ compactification of IIB supergravity to six dimensions, it is important to note that in our supergravity solutions only the volume of $T^4$ is dynamical and thereby we work with  $\cN=1$ supergravity theory in six dimensions without vector multiplets.  This implies that all our supergravity results may be trivially ported to IIB supergravity on $K3$.

\subsection{The IIB solution}
\label{ss:ansatz}

The general solution of type IIB supergravity compactified on $T^4\times S^1$
that preserves the same supercharges as the D1-D5-P system and is
invariant under rotations of $T^4$ has the form
\cite[Appendix E.7]{Giusto:2013rxa}:
\begin{subequations}\label{ansatzSummary}
\allowdisplaybreaks
 \begin{align}
d s^2_{10} &= \frac{1}{\sqrt{\alpha}} \,ds^2_6 +\sqrt{\frac{Z_1}{Z_2}}\,d \hat{s}^2_{4}\, ,\label{10dmetric}\\
d s^2_{6} &=-\frac{2}{\sqrt{\cP}}\,(d v+\beta)\,\Big[d u+\omega + \frac{\mathcal{F}}{2}(d v+\beta)\Big]+\sqrt{\cP}\,d s^2_4\,,\\
e^{2\Phi}&=\frac{Z_1^2}{\cP}\, ,\\
B&= -\frac{Z_4}{\cP}\,(d u+\omega) \wedge(d v+\beta)+ a_4 \wedge  (d v+\beta) + \delta_2\,, \label{Bform}\\ 
C_0&=\frac{Z_4}{Z_1}\, ,\\
C_2 &= -\frac{Z_2}{\cP}\,(d u+\omega) \wedge(d v+\beta)+ a_1 \wedge  (d v+\beta) + \gamma_2\,,\\ 
C_4 &= \frac{Z_4}{Z_2}\, \widehat{\mathrm{vol}}_{4} - \frac{Z_4}{\cP}\,\gamma_2\wedge (d u+\omega) \wedge(d v+\beta)+x_3\wedge(d v + \beta) + {\cal C}\,, \\
C_6 &=\widehat{\mathrm{vol}}_{4} \wedge \left[ -\frac{Z_1}{\cP}\,(d u+\omega) \wedge(d v+\beta)+ a_2 \wedge  (d v+\beta) + \gamma_1\right] \notag\\
  &\hspace{35ex}
  -\frac{Z_4}{\cP}\,{\cal C}\wedge (d u+\omega) \wedge(d v+\beta)\,,
\end{align}
\end{subequations}
with
\begin{equation}
\alpha \equiv \frac{Z_1 Z_2}{Z_1 Z_2 - Z_4^2}~,~~~~
\cP   \equiv     Z_1 \, Z_2  -  Z_4^2 \,.
\label{Psimp}
\end{equation}
Here $ds^2_{10}$ is the ten-dimensional string-frame metric,  $ds^2_{6}$ the six-dimensional Einstein-frame metric, $\Phi$ is the dilaton, $B$ and $C_p$ are the NS-NS and RR gauge forms. (It is useful to also list $C_6$, the 6-form dual to $C_2$, to introduce all the quantities entering the supergravity equations.) The flat metric on $T^4$ is denoted by $d \hat{s}^2_4$ and the corresponding volume form by $\widehat{\mathrm{vol}}_{4}$. The metric  $ds^2_4$ is a generically non-trivial, $v$-dependent Euclidean metric in the four non-compact  directions of the spatial base, $\cB$. We have traded the usual time coordinate, $t$, and the $S^1$ coordinate, $y= x^9$,  for the light-cone coordinates
\begin{equation}
u=\frac{t-y}{\sqrt{2}}\,,\quad v=\frac{t+y}{\sqrt{2}}\,.
\end{equation}
The quantities, $Z_1, Z_2, Z_4, \mathcal{F}$ are scalars;
$\beta,\omega,a_1, a_2, a_4$ are one-forms on $\cB$; $\gamma_1,
\gamma_2, \delta_2$ are two-forms on $\cB$; and $x_3$ is a three-form on
$\cB$.  All these functions and forms can depend not only on the
coordinates of $\cB$ \emph{but also on $v$}.
As discussed below, if the solution is $v$-independent, the one-forms $a_1, a_2,a_4$ may be viewed as five-dimensional Maxwell fields.
Finally, ${\cal C}$ is a
$v$-dependent top form in $\cB$ which can always be set to zero by using
an appropriate gauge~\cite{Giusto:2013rxa}.
To preserve the required supersymmetry, these fields must satisfy
BPS equations \cite{Giusto:2013rxa} and thus get interrelated to one another as we
will explain in subsection \ref{ss:IIBbackground} .

Note that we use the fact that the internal manifold of our solutions is $T^4$ only as an intermediate technical tool, but the final solutions we obtain are solutions of six-dimensional supergravity with two tensor multiplets, which can describe equally well microstate geometries for the D1-D5-P system on K3.

\subsection{The M-theory and five-dimensional pictures}
\label{ss:Mbackground}

Three-charge microstate geometries are expected to be smooth only in the
D1-D5-P duality frame, in which we exclusively work in this paper.
However, it is useful to make connection to other duality frames that
are probably more familiar to the reader, in particular the M-theory
frame in which all the electric charges are  on the same footing and
described by M2-branes.  Moreover, by compactifying M-theory
on $T^6$ and truncating the spectrum one can understand much of the 
structure of the solutions in terms of five-dimensional, $\cN=2$ supergravity 
coupled to $n$ vector multiplets.
However, it is important to note that the M-theory and D1-D5-P frames
are different in one crucial respect:  $v$-dependent solutions in the
D1-D5-P frame, which are essential ingredients of the superstratum
conjecture, are not describable in the M-theory frame, because the
T-duality along the common D1-D5 direction, which connects the two frames, transforms 
$v$-dependent solutions into solutions that contain higher KK harmonics and therefore cannot be described by supergravity.
Therefore, for the purposes of the current paper, the M-theory picture
explained here should be regarded as a book-keeping device to understand
the degrees of freedom appearing in the general three-charge geometries.
We will work in the D1-D5-P frame except for this subsection.

In the five-dimensional description, including the graviphoton, there are thus $(n+1)$
five-dimensional vector fields, $A^{(I)}$, encoded in the eleven-dimensional
three-form potential $C^{(3)}$, and the scalars $t^{(I)}$, encoded in
the K\"ahler form $J$ for the compact six-dimensional space~\cite{Cadavid:1995bk, Papadopoulos:1995da}:
\begin{equation}
C^{(3)}  = \sum_{I = 1}^{n+1} \, A^{(I)} \wedge  J_{I}  ~,~~~
J =  \sum_{I = 1}^{n+1} \, t^{(I)} \wedge  J_{I}  \;.
\label{Cfield}
\end{equation}
Here, $J_{I}$ are harmonic $(1,1)$-forms on the compact six-dimensional space that
are invariant under the projection performing the $\cN=2$ truncation.
In addition the $t^{(I)}$'s satisfy the constraint
\begin{equation}
\coeff{1}{6} \, C_{IJK}  t^{(I)} \, t^{(J)}  \, t^{(k)} ~=~ 1 \,,
\label{tconstraint}
\end{equation}
where $C_{IJK}$ is given by the intersection product among the $J_I$, so
only $n$ scalars are independent.  Here we will take the compact six-dimensional space to
be $T^6$ . If we parametrize the $T^6$ by the holomorphic coordinates:
\begin{equation}
w_1 ~=~ x^5 + i x^6 \,, \qquad   w_2 ~=~ x^7 + i x^8 \,, \qquad   w_3 ~=~ x^9 + i x^{10}  \,,
\label{cplxstr}
\end{equation}
then the requisite forms are  the real and imaginary parts of $d w_a \wedge d \bar w_b$, $a,b =1,2,3$.

5
  
However, we will only need the subset of these:
\begin{align}
 J_1 ~\equiv~  &\coeff{i}{2}\, d w_1 \wedge d \bar w_1 ~=~ dx^5 \wedge  dx^6 \,, \qquad
 J_2 ~\equiv~   \coeff{i}{2}\, d w_2 \wedge d \bar w_2 ~=~ dx^7 \wedge  dx^8  \,, \nonumber \\  
J_3 ~\equiv~ & \coeff{i}{2}\, d w_3 \wedge d \bar w_3 ~=~ dx^9\wedge  dx^{10}  \,, \nonumber\\
J_4~\equiv~ & \coeff{1}{2 \sqrt{2} }\, (d w_1 \wedge d \bar w_2 + d \bar w_1 \wedge d  w_2)  ~=~   \coeff{1}{\sqrt{2} }\, (dx^5 \wedge  dx^7 + dx^6 \wedge  dx^8)     \,,  \nonumber \\
J_5 ~\equiv~ &   \coeff{i}{2\sqrt{2}}\, (d w_1 \wedge d \bar w_2 - d \bar w_1 \wedge d  w_2)   ~=~     \coeff{1}{\sqrt{2} }\, (dx^5 \wedge  dx^8 - dx^6 \wedge  dx^7)  \,.
\label{harmforms}
\end{align}
In this basis, the only non-zero components of $C_{IJK}$ are   
\begin{equation}
C_{3JK}  ~\equiv~  \widehat C_{JK}  ~=~  \widehat C_{JK} \,, \qquad J, K \in \{1,2,4,5\}
\label{RecC}
\end{equation}
where
\begin{equation}
\widehat C_{JK}  ~=~  
\begin{pmatrix} 
0&1&0&0\\  1&0&0&0\\ 0&0&-1&0\\ 0&0&0&-1
\end{pmatrix}   \,.
\label{hatCform}
\end{equation}
The standard ``STU'' supergravity corresponds to setting $A^{(5)} = A^{(4)} = 0$ and retaining $A^{(I)}$, $I =1,2,3$, with $A^{(1)}+A^{(2)}+A^{(3)}$ being  the graviphoton. The degrees of freedom in our particular examples of a superstratum correspond to the presence of one extra vector multiplet and this involves identifying  in the fourth and fifth sets of fields: $A^{(5)} = A^{(4)}$ and $t^{(5)}=t^{(4)}$ as in \cite{Giusto:2012gt,Vasilakis:2012zg}.

 In the ``STU'' model, the standard route (see, for example, the appendices in
\cite{Bena:2008dw}) for getting from  the IIB frame to the M-theory
frame is to perform T-dualities on $(x^9, x^5, x^6)$ and then to uplift
the resulting IIA description to eleven dimensions. In the IIB solution there are four
independent scalar functions ($\mathcal{F}\equiv -2 (Z_3-1)$ and $Z_I$
with $I=1,2,4$) whose ratios correspond to the scalars in the vector
multiplets.   The fourth function represents  a convenient way of writing the 
warp factor of the five-dimensional metric as a relaxation of the constraint (\ref{tconstraint}):
\begin{equation}
Z ~\equiv~  \big(\coeff{1}{6} \, C^{IJK} Z_I \, Z_J \, Z_K\big)^{\frac{1}{3}}\,,
\label{Zreln}
\end{equation}
In our particular class of solutions with  (\ref{RecC}) and (\ref{hatCform}) we have
\begin{equation}
Z^3   ~=~  \coeff{1}{2} \, Z_3 \, \big( \widehat C^{IJ}   Z_I \, Z_J \big)~=~   Z_3 \, \big(   Z_1 \, Z_2  - Z_4^2  \big) ~=~ Z_3 \cP\,,
\label{Psimp2}
\end{equation}
where $\widehat C^{IJ}\equiv \widehat C_{IJ} $ and  we identified $A^{(5)} = A^{(4)}$.  
The combination $\cP$ will be ubiquitous as a warp factor in the six-dimensional formulation. The function, ${\cal F}$, and the vector field, $\beta$, encode the momentum and the KK-monopole charges and form the time and the space components of the five-dimensional  vector $A^{(3)}$, while the other two scalars $Z_1$ and $Z_2$ combine with $a_1$ and $a_2$ to give  $A^{(1)}$ and $A^{(2)}$. As mentioned above, the degrees of freedom of the IIB solution~\eqref{ansatzSummary} require an extra vector multiplet. In order to map the IIB configuration in the M-theory frame one needs a slightly more complicated combination of T-dualities and one S-duality~\cite{Giusto:2011fy}. However the final result is very similar to that of the ``STU'' model, with the scalar $Z_4$ and vector $a_4$ forming the new vector multiplet. 

One can also uplift the five-dimensional description given here to  $\Neql1$ supergravity in six dimensions.~\cite{Romans:1986er,Ferrara:1997gh}. This is more appropriate for our solution, since this formulation allows $v$-dependent solutions. Indeed, the six-dimensional formulation is the $T^4$ reduction of the IIB description in Section \ref{ss:ansatz}.  In the uplift, the five-dimensional graviton multiplet combines with one of the vector multiplets to yield the six-dimensional graviton multiplet, while all the remaining vector multiplets become anti-self-dual tensor multiplets.  Thus the ``STU'' model corresponds to minimal $\Neql1$ supergravity (whose bosonic sector consists of a graviton, $g_{\mu \nu}$, and a self-dual tensor gauge field, $B^+_{\mu \nu}$) plus a tensor multiplet (whose bosonic sector consists of an anti-self-dual tensor gauge field, $B^-_{\mu \nu}$, and a scalar, $\Phi$). The BPS equations for these systems were obtained in \cite{Gutowski:2003rg,Cariglia:2004kk} and were fully analyzed and greatly simplified in \cite{Bena:2011dd}. To build our solutions we need to add an extra anti-self-dual tensor multiplet and the corresponding analysis of the BPS equations is discussed in~\cite{Giusto:2013rxa}. We now summarize this result and present the equations that we need to solve in order to construct a superstratum in the class of solutions presented in~\eqref{ansatzSummary}.

\subsection{The equations governing the supersymmetric solutions}
\label{ss:IIBbackground}

The BPS conditions require that everything be $u$-independent and, in
particular, $\frac{\partial}{\partial u}$ must be a Killing vector. It is convenient to think of the fields in terms of the
four-dimensional base geometry and so one defines a covariant exterior
derivative
\begin{equation}
\mathcal{D} \equiv d - \beta\wedge \frac{\partial}{\partial v}\,.
\end{equation}
Here and throughout the rest of the paper, $d$ denotes the exterior
differential on the spatial base $\cB$ \footnote{Note that this convention differs from that of much of the earlier literature in which the exterior differential on the spatial base $\cB$ is denoted by $\tilde d$.}. The
derivative, $\mathcal{D}$, is covariant under diffeomorphisms mixing $v$
and $x^i$:
\begin{equation}
v\to v - V(x^i)\,,\quad \beta \to \beta + dV\,.
\end{equation}
Given that everything is $u$-independent, the class of diffeomorphisms of $u$ that respect the form of the solution
\eqref{ansatzSummary}
may be recast in terms of a gauge invariance:
\begin{equation}\label{gaugecalF}
u\to u + U(x^i,v)\,,\quad \omega \to \omega - dU + \dot{U}\,\beta \,,\quad \mathcal{F} \to \mathcal{F} - 2\, \dot{U}\,,
\end{equation}
where a dot denotes differentiation with respect to $v$.

It was shown in \cite{Bena:2011dd} that the supersymmetry constraints
and the equations of motion have a linear structure and this will be crucial for
the construction of solutions. The only intrinsically non-linear subset
of constraints (the ``zeroth layer'' of the problem) is the one that involves the four-dimensional metric,
$ds^2_4$, and the one-form $\beta$. In this paper we restrict to a class
of solutions where these constraints are trivially satisfied: we take
the spatial base $\cB$ to be $\mathbb{R}^4$ and its metric $ds^2_4$ to
be the flat, $v$-independent metric. We will also require $\beta$ to be
$v$-independent \emph{but all other functions and fields will be allowed
to be $v$-dependent.}   In this situation, the BPS equations for $\beta$ now 
reduce to the simple, linear requirement that  $\beta$ has self-dual field strength
 \begin{equation}\label{eqbeta}
d \beta = *_4 d\beta\,,
 \end{equation}
where $*_4$ denotes the flat $\mathbb{R}^4$ Hodge dual. 

We have described the solution in terms of gauge potentials $B$ and
$C_p$ ($p=0,2,4,6$) but this means that some of the fields do not have a
gauge invariant meaning. The field strengths can be written
\cite{Giusto:2013rxa} in terms of the 2-forms
\begin{equation}
\label{Thetadefs}
\Theta_1 \equiv \mathcal{D} a_1 + \dot{\gamma_2}\,,\quad \Theta_2 \equiv \mathcal{D} a_2 + \dot{\gamma_1}\,,\quad \Theta_4 \equiv \mathcal{D} a_4 + \dot{\delta_2}\,,
\end{equation}
and the 4-form, $\Xi_4 $, obtained from $x_3$ 
\begin{equation}\label{eq:Xi4}
\Xi_4 =  \mathcal{D} x_3 - \Theta_4\wedge \gamma_2+ a_1 \wedge (\mathcal{D} \delta_2- a_4\wedge d \beta) + {\cal C} \,.
\end{equation}
The combinations in~\eqref{Thetadefs} are invariant under the transformations $a_1 \to a_1 - \dot{\xi}$, $\gamma_2 \to \gamma_2 + {\cal D} \xi$ where $\xi$ is a 1-form, and similarly for $(a_2, \gamma_1)$ and $(a_4,\delta_2)$. The 4-form in~\eqref{eq:Xi4} is invariant under the transformation involving $a_1$ provided that $x_3 \to x_3 + \Theta_4 \wedge \xi$ and ${\cal C} \to {\cal C} - *_4 {\cal D} Z_4 \wedge \xi$, as it can be checked by using~\eqref{eqZ4Theta4}.
 
The next layer (the ``first layer'') of BPS equations determine the warp factors $Z_1$, $Z_2$,
$Z_4$ and the gauge 2-forms $\Theta_1$, $\Theta_2$,
$\Theta_4$:\footnote{Using the intersection numbers 
\eqref{hatCform}, the equations
\eqref{eqZ1Theta2}--\eqref{eqcalF} can be written more succinctly as
\begin{equation}
\begin{gathered}
  *_4\cD \dot{Z}'_I=\widehat{C}_{IJ}\cD \Theta^J,\quad
 \cD*_4\cD Z'_I=-\widehat{C}_{IJ}\Theta^I\wedge d\beta,\quad
 \Theta^I=*_4\Theta^I,\qquad
 (1+*_4)\cD \omega + \cF \,d\beta  = Z'_I \Theta^I,\\
 *_4\mathcal{D} *_4\!\Bigl(\dot{\omega} -\frac{1}{2}\,\mathcal{D} \mathcal{F}\Bigr)
 =
 \half\partial_v(\widehat{C}^{IJ}Z'_I Z'_J) 
 -\half\widehat{C}^{IJ}\dot{Z}'_I \dot{Z}'_J
 -{1\over 4}\widehat{C}_{IJ}*_4(\Theta^I\wedge \Theta^J),
\end{gathered}
\end{equation}
where $Z'_1\equiv Z_1$, $Z'_2\equiv Z_2$,
$Z'_4=Z'_5\equiv-Z_4$, $\Theta_4=\Theta_5$, and
$\Theta^I\equiv\Theta_I$.  }
 \begin{equation}\label{eqZ1Theta2}
 *_4 \mathcal{D} \dot{Z}_1 =  \mathcal{D} \Theta_2\,,\quad \mathcal{D}*_4\mathcal{D}Z_1 = -\Theta_2\wedge d\beta\,,\quad \Theta_2=*_4 \Theta_2\,,
 \end{equation}
  \begin{equation}\label{eqZ2Theta1}
 *_4 \mathcal{D} \dot{Z}_2 =  \mathcal{D} \Theta_1\,,\quad \mathcal{D}*_4\mathcal{D}Z_2 = -\Theta_1\wedge d\beta\,,\quad \Theta_1=*_4 \Theta_1\,,
  \end{equation}
  \begin{equation}\label{eqZ4Theta4}
 *_4 \mathcal{D} \dot{Z}_4 =  \mathcal{D}  \Theta_4\,,\quad \mathcal{D}*_4\mathcal{D}Z_4 = -\Theta_4\wedge d\beta\,,\quad \Theta_4=*_4 \Theta_4\,.
\end{equation}
It is worth noting that the first equation in each set involves four component equations, while the second  equation in each set is essentially an integrability condition for the first equation.  The self-duality condition reduces each $ \Theta_j$ to three independent components and adding in the corresponding $Z_k$ yields four independent functional components upon which there are four constraints. 

The final layer (the ``second layer'') of constraints are linear equations for $\omega$ and $\mathcal{F}$:
 \begin{equation}\label{eqomega}
\mathcal{D} \omega + *_4  \mathcal{D}\omega + \mathcal{F} \,d\beta = Z_1 \Theta_1+ Z_2 \Theta_2 -2\,Z_4 \Theta_4 \,,
\end{equation}
and a second-order constraint that follows from the $vv$ component of Einstein's equations\footnote{This simplified form is completely equivalent to (2.9b) of~\cite{Giusto:2013bda}.}
\begin{equation}\label{eqcalF}
\begin{aligned}
 *_4\mathcal{D} *_4\!\Bigl(\dot{\omega} -\frac{1}{2}\,\mathcal{D} \mathcal{F}\Bigr)&~=~\dot{Z}_1\dot{Z}_2+Z_1 \ddot{Z}_2 + Z_2 \ddot{Z}_1 -(\dot{Z}_4)^2 -2 Z_4 \ddot{Z}_4-\frac{1}{2} *_4\!\Big(\Theta_1\wedge \Theta_2 - \Theta_4 \wedge \Theta_4\Bigr) \\
 &~=~\partial_v^2 (Z_1 Z_2 - {Z}_4^2)  -(\dot{Z}_1\dot{Z}_2  -(\dot{Z}_4)^2 )-\frac{1}{2} *_4\!\Big(\Theta_1\wedge \Theta_2 - \Theta_4 \wedge \Theta_4\Bigr)\,.
\end{aligned}
\end{equation}
The important point is that these equations determine the complete solution and form a system that can be solved in a linear sequence, because the right-hand side of each equation is made of source terms that have been computed in the preceding layers of the BPS system.

\subsection{Outline of the construction of a superstratum}
\label{ss:Goal}

We start in much the same way as in \cite{Bena:2011uw,Niehoff:2013kia,Bena:2014qxa}, with a round, D1-D5 supertube solution, in the decoupling limit.  The  geometry of this background is \emph{global} AdS$_3$ $\times S^3$.  The $SU(2)_L \times SU(2)_R$ isometry of the $S^3$ corresponds to the $\cR$-symmetry and the $SL(2,\mathbb{R})_L \times SL(2,\mathbb{R})_R$ isometry of the AdS$_3$ yield the finite left-moving and right-moving conformal groups.  The mode analysis and holographic dictionary of this background is extremely well-understood \cite{Kanitscheider:2006zf,Kanitscheider:2007wq}. The background is dual to the Ramond ground state with maximal angular momentum: $j=\jt = (n_1 n_5)/2$, $h=\tilde h =0$, with $j$ the eigenvalue of the $SU(2)_L$ generator $J^3_0$, and $h$ the eigenvalue of the $SL(2,\mathbb{R})_L$ generator $L_0 - c/24$ (tilded quantities denote the right-moving sector counterparts).  $n_1$ and $n_5$ are the number of D1 and D5-branes, respectively. The ``supertube'' shape modes associated with generic $\frac{1}{4}$-BPS  D1-D5 states have $j, \jt\le (n_1 n_5)/2$,  but always $h=\tilde h=0$. In particular,  $|j-\jt|$ is the spin of the underlying supergravity field. Thus,  for a fixed spin field, these shape Fourier modes are determined by one quantum number and hence correspond to one-dimensional shape modes.  

Adding momentum modes while maintaining $\frac{1}{8}$-supersymmetry means that we allow more general excitations in the left sector of the CFT in such a way that $h>0$, while  preserving the right-sector structure of the excitation (and hence $\tilde h=0$). Thus, generic $\frac{1}{8}$-BPS modes will have quantum numbers $(j, h; \jt, \tilde h=0)$. Since  $h$ is independent of $j$, these will generate intrinsically two-dimensional shape modes, for fixed spin. In this way, we can think of the superstratum as two-dimensional shape modes on the homology $3$-cycle of the underlying microstate geometry.

It is also useful to consider the NS sector states obtained by spectral flow from the Ramond sector.  Ramond ground states are mapped to chiral primaries, which have $j=h$ and $\jt=~\tilde h$. Acting on chiral primaries with $SU(2)_L \times SL(2,\mathbb{R})_L$ generators generically gives non-chiral primaries with $j\not = h$, which map back to states carrying momentum in the Ramond sector\cite{Mathur:2003hj}. 

Since we know the action of $SU(2)_L \times SL(2,\mathbb{R})_L$ on
gravity fields, we can construct the modes corresponding to
descendants\footnote{The states obtained by acting $\cR$-symmetry
generators on a chiral primary state must more precisely be called
super-descendants, but for simplicity we refer to them as descendants.}
of chiral primaries \cite{Giusto:2013bda}. At the linearized level, we
can take arbitrary linear combinations of these modes to make the
superstratum. As we will see more explicitly in the next section, this
will give us the solution of the first layer of the BPS equations. To
construct the fully non-linear solution, we use the power of the
observation \cite{Bena:2011dd} that the upper layers of the BPS
equations are a linear system of equations. This means that the linear
excitations can be used directly to obtain the complete solution in
which the fluctuations are \emph{large}.  While simple, in principle,
there are several essential technical obstacles to be overcome:
\begin{itemize}
\item[(i)]  The construction of the generic linear modes explicitly in some manageable form. 
\item[(ii)] Solving the linear equations for the upper layers \emph{with sources} constructed from combinations of the linearized modes.
\item[(iii)] Removal of the singularities and building a smooth solution by fixing some of the Fourier modes but doing so in a manner that leaves two arbitrary quantum numbers, thus preserving the intrinsically two-dimensional form of the fluctuations.   
\end{itemize}
We now proceed to solve each of these problems one after another.  This will mean that we have to dive into some very technical computations but we will regularly step back and orient the reader in terms of the goals stated here.

\section{Solving the first layer of BPS equations}
\label{Sect:layer1}

While supersymmetry does not allow the solutions to depend on $u$, states carrying momentum are generically going to be $v$-dependent. In the rest of this paper we will make the simplifying assumption that the four-dimensional metric $ds^2_4$ is $v$-independent and simply that of flat $\mathbb{R}^4$.  We also assume  that the one-form $\beta$, which determines the KKM fibration along the D1-D5 common direction, is  $v$-independent.  We make this assumption simply for  expediency; we do not know how to solve the system otherwise.  These assumptions could, in principle,  prevent us from finding a  ``suitably generic'' superstratum because all the fluctuations that we will introduce in the other fields may ultimately require $v$-dependent base metrics and $v$-dependent $\beta$ in order for the solution to be smooth. Indeed, generic superstrata will have $v$-dependence everywhere but our goal here is to demonstrate that there is at least  one class of superstrata that is a ``suitably generic''  function of two variables.  The fact that we will succeed despite this technical restriction is remarkable even though there are {\it a posteriori\/} explanations of this somewhat miraculous outcome.

\subsection{Two-charge solutions}
\label{ss:seed}

It is useful to think of the three-charge solutions as obtained by adding momentum-carrying perturbations to some two-charge seed. This will not only facilitate the CFT interpretation of the states but also give important clues for the construction of the geometries. All two-charge D1-D5 microstates have been constructed in \cite{ Lunin:2001jy, Lunin:2002bj,Lunin:2002iz,  Kanitscheider:2007wq} and are associated with a closed curve in $\mathbb{R}^8$, $g_A(v')$ ($A=1,\dots,8$). This curve has the interpretation of the profile of the oscillating fundamental string dual to the D1-D5 system.  The parameter along the curve is $v'$, which has a  periodicity  $L= 2\pi \frac{Q_5}{R}$ where $Q_5$ is the D5 charge and $R$ is the radius of $S^1$.

In the duality frame of the fundamental string, the profile can be split
into four $\mathbb{R}^4$ components ($A=1,\dots,4$) and four $T^4$
components ($A=5,\dots,8$).  The states with non-vanishing $g_A(v')$ for
$A=5,\dots,8$ break the symmetry of $T^4$ but, when one dualizes to the
D1-D5 duality frame, one of the $T^4$ components, which we take to be
$A=5$, plays a distinct role and, in fact, the D1-D5 geometries that
have non-trivial values of $g_A(v')$ for $A=1,\dots,5$ are invariant
under rotations of $T^4$.  These solutions therefore fall in the class
described by the class of solutions \eqref{ansatzSummary}. We recall in
Appendix~\ref{Appendix:2charge} how to generate the geometry from the
profile $g_A(v')$ for this restricted class of two-charge states.

 The simplest two-charge geometry is that of a round supertube, described by a circular profile in the $(1,2)$ plane:
 \begin{equation}\label{circle}
 g_1(v') = a \cos\Bigl(\frac{2\pi\,v'}{L}\Bigr)\,,\quad g_2(v') =  a \sin\Bigl(\frac{2\pi\,v'}{L}\Bigr)\,,\quad g_A(v')=0\quad \mathrm{for}\quad A=3,\ldots,8\,.
 \end{equation}
 The metric of the supertube is more easily expressed in the spheroidal, or two-centered, coordinates in which $\mathbb{R}^4$ is parameterized as
 \begin{equation}
 x_1 + i x_2 = \sqrt{r^2+a^2}\,\sin\theta\,e^{i \phi}\,,\quad x_3 + i x_4 = r \,\cos\theta\,e^{i \psi}\,.
 \end{equation}
The locus $r=0$ thus describes a disk of radius $a$ parameterized by $\theta$ and $\phi$ with the origin of  $\mathbb{R}^4$  at $(r=0,\theta=0)$ while the tube  lies at the perimeter of this disk $(r=0,\theta=\pi/2)$.    In these coordinates the flat $\mathbb{R}^4$ metric is
 \begin{equation}\label{ds4flat}
 d s^2_4 = (r^2+a^2 \cos^2\theta)\Bigl(\frac{d r^2}{r^2+a^2}+ d\theta^2\Bigr)+(r^2+a^2)\sin^2\theta\,d\phi^2+r^2 \cos^2\theta\,d\psi^2\,.
\end{equation}
The metric coefficients specifying the supertube geometry are
\begin{subequations}\label{supertube}
\allowdisplaybreaks
\begin{equation}
Z_1 = 1+\frac{Q_1}{\Sigma}\,,\quad Z_2 = 1+\frac{Q_5}{\Sigma}\,, \\
\end{equation}
\begin{equation}
\beta =  \frac{R\,a^2}{\sqrt{2}\,\Sigma}\,(\sin^2\theta\, d\phi - \cos^2\theta\,d\psi)\,,\quad \omega=\frac{R\,a^2}{\sqrt{2}\,\Sigma}\,(\sin^2\theta\, d\phi + \cos^2\theta\,d\psi)\,,
\end{equation}
\begin{equation}
Z_4=\mathcal{F}=0\,,\quad \Theta_1=\Theta_2=\Theta_4=0\,,
\end{equation}
\end{subequations}
where
\begin{equation}
\Sigma \equiv r^2 + a^2 \cos^2\theta\,.
\end{equation}
The parameter $a$ is related to the D1 and D5 charges $Q_1$, $Q_5$ and the radius, $R$, of $S^1$ by
\begin{equation}
R = \sqrt{\frac{Q_1 Q_5}{a}}\,.
\end{equation}
As one would expect, this geometry is asymptotic to $\mathbb{R}^{4,1}
\times S^1 \times T^4$.  The charges $Q_1$ and $Q_5$ are related to the
quantized D1, D5-brane numbers, $n_1$ and $n_5$, by the relation \eqref{Q1Q5_n1n5}.

\subsection{The solution generating technique}
\label{ss:newmodes}

As usual, one can define a decoupling limit which corresponds to cutting off the asymptotic part of the geometry.  This is achieved by taking
\begin{equation}
r \ll \sqrt{Q_i} \ll R \quad (i=1,5)\,,
\end{equation}
and it implies that the ``1'' in the warp factors $Z_1$ and $Z_2$ can be neglected. In this limit, the supertube geometry reduces to AdS$_3\times S^3 \times T^4$, as one can  explicitly verify by performing the  coordinate redefinition 
\begin{equation}\label{spectralflow}
\phi\to \phi +\frac{t}{R}\,, \quad \psi\to \psi +\frac{y}{R}\,
\end{equation}
in the geometry \eqref{10dmetric} with the data \eqref{supertube}.

Working in the decoupling region
has the advantage that one can generate new solutions via the action of
the symmetries of the CFT\@.  These symmetries form a chiral algebra whose rigid limit
is $SU(2)_L\times SU(2)_R\times SL(2,\mathbb{R})_L \times
SL(2,\mathbb{R})_R\times U(1)^4$. On the gravity side, each CFT
transformation is realized by a diffeomorphism that is non-trivial at
the AdS boundary. The $SU(2)$ factors are $\cR$-symmetries of the CFT
with generators $\{J^i_0,\Jt^i_0\}$, $i=\{\pm,3\}$, and correspond, in
gravity, to rotations of $S^3$.  The $SL(2,\mathbb{R})$ factors, with
generators $\{L_0,L_{\pm 1},\tilde{L}_0,\tilde{L}_{\pm 1}\}$, are
conformal transformations in AdS$_3$.  The $U(1)$ factors are torus
translations. The extension of these transformations to the full chiral
algebra with $\{J^{i}_{-n},\Jt^i_{-n},L_{-n},\tilde{L}_{-n}\}$,
$n\in\bbZ$ is discussed, from the gravity point of view, in
\cite{Mathur:2011gz}. The affine extension of $U(1)$ torus translations was considered in \cite{Lunin:2012gp} and used to generate an exact family of three-charge solutions.  

One can generate a three-charge solution by acting on a two-charge solution by a generator with $n\ge 1$, because the level, $n$, corresponds to the third (momentum) charge\footnote{If we take the decoupling limit of the two-charge solution, the corresponding state in the boundary CFT is a ground state in the RR sector.  By the ``level'' here, we mean the one in the RR sector. The momentum charge $n_p$ is given by $n_p=L_0^{\rm RR}-\tilde{L}_0^{\rm RR}$. If one excites the left-moving sector only, this gives $n_p=L_0^{\rm RR}$ (modulo the zero-point energy shift by $-c/24$).}.
To preserve half (four supercharges) of the supersymmetry preserved by the two-charge state, one can only act with generators in the left-moving sector. For example, one can consider the transformation $e^{\chi (J^+_{-1}- J^-_1)}$, whose action on a particular two-charge state was studied in \cite{Giusto:2013bda}, while the action on generic two-charge states at the linearized level was found in \cite{Shigemori:2013lta}. The action of this operator is particularly easy to implement, because  $J^+_{-1}- J^-_1$ is related to the rotation, $J^+_{0}- J^-_0=2iJ^2_0$, on $S^3$ by the change of coordinates that generates a spectral flow (\ref{spectralflow})\cite{Mathur:2003hj}.  Explicitly, the relation is
 \begin{equation}
 J^+_{-1}- J^-_1 = e^S (J^+_{0}- J^-_0) e^{-S}\,,
 \end{equation}
where $e^{-S}$ describes the coordinate transformation (\ref{spectralflow}).

The simplest two-charge geometry corresponding to the round profile \eqref{circle} is mapped by this $e^{-S}$ coordinate transformation to the space AdS$_3\times S^3 \times T^4$, which is rotationally invariant. Therefore, the operator $e^{\chi (J^+_{-1}- J^-_1)}$ acts trivially on the round supertube seed solution and we do not get a new three-charge solution.
In order to generate a non-trivial three-charge solution, instead, one should start with a deformed two-charge seed. 

\subsection{A ``rigidly-generated'' three-charge solution}
\label{ss:particularSoln}

Perhaps the simplest two-charge seed solution\footnote{Another possibility is to turn on a ``density fluctuation'' on the profile (\ref{circle}) by changing the profile parametrization as $v'\to \Lambda(v')$ for some function $\Lambda(v')$; the corresponding geometry would have undeformed 1-forms and no $Z_4$ would be generated, but $Z_1$ and $Z_2$ would be modified.} that can be used to generate a new three-charge solution is the one obtained by turning on the $A=5$ component of the profile $g_A$. This produces a three-charge geometry that fits in the class (\ref{ansatzSummary}), has undeformed one-forms $\beta$ and $\omega$, but a non-trivial $Z_4$.  Concretely, we consider the following profile as the seed:
 \begin{equation}\label{circleplus}
 g_1(v') = a \cos\Bigl(\frac{2\pi\,v'}{L}\Bigr)\,,\quad g_2(v') =  a \sin\Bigl(\frac{2\pi\,v'}{L}\Bigr)\,,\quad g_5(v')=-\frac{b}{k} \sin\Bigl(\frac{2\pi\,k\,v'}{L}\Bigr)\,,
 \end{equation}
where $k$ is a positive integer and the remaining components of $g_A$ remain trivial. The corresponding two-charge geometry is described by
 \begin{subequations}\label{supertubeplus}
\allowdisplaybreaks
\begin{equation}
Z_1 = \frac{R^2}{2\,Q_5}\Bigl[\frac{2a^2+b^2}{\Sigma} + \,b^2\,a^{2k}\,\frac{\sin^{2k}\theta\,\cos(2k\phi)}{(r^2+a^2)^k\,\Sigma}\Bigr]\,,\quad Z_2 = \frac{Q_5}{\Sigma}\,, \\
\end{equation}
\begin{equation}\label{eq:bo2c}
\beta =  \frac{R\,a^2}{\sqrt{2}\,\Sigma}\,(\sin^2\theta\, d\phi - \cos^2\theta\,d\psi)\,,\quad \omega=\frac{R\,a^2}{\sqrt{2}\,\Sigma}\,(\sin^2\theta\, d\phi + \cos^2\theta\,d\psi)\,,
\end{equation}
\begin{equation}
Z_4=R\,b\,a^k\,\frac{\sin^k\theta\,\cos(k\phi)}{(r^2+a^2)^{k/2}\,\Sigma}\,,
\end{equation}
\begin{equation}
\mathcal{F}=0\,,\quad \Theta_1=\Theta_2=\Theta_4=0\,,
\end{equation}
\end{subequations}
where we are restricting to the decoupling region and hence have dropped the
``$1$'' in $Z_1$ and $Z_2$. The relation between the parameters $a$,
$b$, the asymptotic charges $Q_1,Q_5$, and the $S^1$ radius $R$ is now
\begin{equation}\label{eq:abq1q5}
R =\sqrt{ \frac{Q_1 Q_5}{a^2+\frac{b^2}{2}}}\,.
\end{equation}
For fixed $Q_1$, $Q_5$, $R$ and $k$, the solutions thus admit a freely varying parameter, that could be taken to be $b/a$. We will discuss in Section~\ref{Sect:D1D5CFT} the CFT interpretation of this family of two-charge solutions. 

We note the appearance of a non-trivial, $\phi$-dependent $Z_4$, which is accompanied by a $\phi$-dependent deformation, at second order in the deformation parameter $b$, in $Z_1$.  The function  $Z_2$ remains unchanged. It is also very interesting to note that the combination $Z_1 Z_2 - Z_4^2$ is deformed at order $b^2$, but the form of the $\phi$-dependent terms in $Z_4$ and $Z_1$ is such  that $Z_1 Z_2 - Z_4^2$ is $\phi$-independent.  As a result, the six-dimensional Einstein metric does not depend on $\phi$.  This is very similar to the mechanism that plays a central role in obtaining neutral black hole microstate geometries \cite{Mathur:2013nja} and smooth ``coiffured'' black rings \cite{Bena:2014rea}.

Now we apply the solution generating technique by acting with $e^{\chi
(J^+_{-1}- J^-_1)}$ on the two-charge solutions \eqref{supertubeplus}
with nonzero $b$  and obtaining a new three-charge
solution\footnote{The explicit change of coordinates realizing $e^{\chi
(J^+_{-1}- J^-_1)}$ on the gravity side can be found in
\cite{Giusto:2013bda}.} \cite{Giusto:2013bda}.  The resulting solution
represents a very particular three-charge state which, by construction,
is a chiral algebra descendant of a two-charge state.
We will refer to such a  solution as a ``rigidly-generated'' three-charge solution but we will use this solution as an inspiration to construct far more general classes of solution 
that are far from being rigid, and, in particular, are no longer descendants of two-charge states. 
It turns out that the transformation $e^{\chi (J^+_{-1}- J^-_1)}$ does
not modify the four-dimensional metric and the one-form $\beta$. Namely,
our particular rigidly-generated three-charge solution still has
\begin{equation}
 d s^2_4 = \Sigma \Bigl(\frac{d r^2}{r^2+a^2}+ d\theta^2\Bigr)+(r^2+a^2)\sin^2\theta\,d\phi^2+r^2 \cos^2\theta\,d\psi^2\,,
\label{ds24baseUnchg}
 \end{equation}
and
\begin{equation}
\beta =  \frac{R\,a^2}{\sqrt{2}\,\Sigma}\,(\sin^2\theta\, d\phi - \cos^2\theta\,d\psi)\,.
\label{betaUnchg}
\end{equation}
As mentioned in Section~\ref{ss:IIBbackground}, we assume that the same happens in all three-charge geometries we consider, even if they are not descendant of two-charge microstates.  So, hereafter, we always assume that $ds^2_4$ and $\beta$ are given by \eqref{ds24baseUnchg} and \eqref{betaUnchg}.

The $Z_4$ in the  rigidly-generated solution is a linear superposition of modes of the form \cite{Giusto:2013bda}:
\begin{equation}
Z_4^{(k,m)} = R\,\frac{\Delta_{k,m}}{\Sigma} \cos\biggl(
m \frac{\sqrt{2}\,v}{R}+(k-m)\,\phi - m\,\psi\biggr) \label{Z4modes}\,,  
\end{equation}
with
\begin{equation}\label{eq:Deltakm}
\Delta_{k,m}\equiv \Bigl(\frac{a}{\sqrt{r^2+a^2}}\Bigr)^k\,\sin^{k-m}\theta\,\cos^m\theta\,.
\end{equation}
The solution also has a non-trivial spatial component of the NS-NS
$2$-form in (\ref{Bform}). One finds that this may be most simply
written in terms of the gauge invariant quantities
 \begin{align}\label{Theta4modes}
\Theta_4^{(k,m)} = & -\sqrt{2} \, m\,\Delta_{k,m} \,
r\sin\theta\,\,\Omega^{(1)}\, \sin\biggl(m \frac{\sqrt{2}\,v}{R}+(k-m)\,\phi - m\,\psi\biggr) 
\\ \nonumber
&  -\sqrt{2} \, m\,\Delta_{k,m}\, \Omega^{(2)}\, \cos\biggl(
m \frac{\sqrt{2}\,v}{R}+(k-m)\,\phi - m\,\psi\biggr)\,,
\end{align}
where $\Omega^{(1)}$, $\Omega^{(2)}$ and $\Omega^{(3)}$ are a basis of self-dual 2-forms on $\mathbb{R}^4$:
\begin{equation}\label{selfdualbasis}
\begin{aligned}
\Omega^{(1)} &\equiv \frac{dr\wedge d\theta}{(r^2+a^2)\cos\theta} + \frac{r\sin\theta}{\Sigma} d\phi\wedge d\psi\,,\\
\Omega^{(2)} &\equiv  \frac{r}{r^2+a^2} dr\wedge d\psi + \tan\theta\, d\theta\wedge d\phi\,,\\
 \Omega^{(3)} &\equiv \frac{dr\wedge d\phi}{r} - \cot\theta\, d\theta\wedge d\psi\,.
\end{aligned}
\end{equation}
Note that these are \emph{not} normalized but satisfy
\begin{equation}
\begin{aligned}
  *_4(\Omega^{(1)}\wedge \Omega^{(1)})
 &={2\over (r^2+a^2)\Sigma^2\cos^2\theta},\quad&
 *_4(\Omega^{(2)}\wedge \Omega^{(2)})
 &={2\over (r^2+a^2)\Sigma\cos^2\theta},\\
 *_4(\Omega^{(3)}\wedge \Omega^{(3)})
 &={2\over r^2\Sigma\sin^2\theta},&
 *_4(\Omega^{(i)}\wedge \Omega^{(j)})&=0,\quad i\neq j.
\end{aligned}
\end{equation}

For generic values of the rotation angle $\chi$, one finds that all terms with $m\le k$ appear in the rigidly-generated solution. We will see in Section~\ref{Sect:D1D5CFT} that this happens because the operator $(J^+_{-1})^m$ annihilates the two-charge state if $m>k$. The reflection of this fact on the gravity side is that the functions $\Delta_{k,m}$ are obviously singular for $\theta=0$ if $m>k$ and thus should not appear in physically allowed solutions. Hence the modes, $Z_4^{(k,m)}$ and $\Theta_4^{(k,m)}$, are only allowed if $m\le k$.  Note that, for these modes, the functions multiplying the $v,\phi,\psi$-dependent trigonometric functions vanish fast enough to avoid singularities at $(r=0,\theta=0)$ according to the criterion discussed at the beginning of Section~\ref{Sect:layer2}.
In the rigidly-generated solution, the coefficients with which the terms $Z_4^{(k,m)}$ appear in the total $Z_4$ are not all independent, but are fixed functions of a single parameter, the rotation angle $\chi$.\footnote{For an explicit example see Appendix A of \cite{Giusto:2013bda}. From that example one can also see that there exists one particular value of $\chi$ ($\chi=\pi/2$) for which all coefficients apart from those of the terms with $m=k$ vanish: this shows that solutions where $Z_4$ contains only modes with $m=k$ are descendants of two-charge states.}

\subsection{A general class of solutions to the first layer}
\label{ss:genfirstlayer}

The beauty of the solution generating technique is that it provides us
with all the modes we need to solve the first layer of the BPS
equations; indeed, one can explicitly check that \emph{each individual
mode} given by \eqref{Z4modes} and \eqref{Theta4modes} solves the first
layer of equations, \eqref{eqZ4Theta4}.  These modes depend upon two
integers, $(k,m)$, and provide an expansion basis for generic functions
of two variables on the $S^3$.  So, as far as this layer of the problem
is concerned, we can take advantage of the linearity of the BPS system
and consider solutions in which $Z_4,\Theta_4$ are linear combinations of
$Z_4^{(k,m)}$, $\Theta_4^{(k,m)}$ with \emph{arbitrary} coefficients:
\begin{equation}\label{Z4full}
Z_4 = R\,\sum_{(k,m)} b_{k,m}\, \frac{\Delta_{k,m}}{\Sigma} \,\cos\hat{v}_{k,m} \,,
\end{equation}
\begin{equation}
\label{Theta4full}
\Theta_4 = -\sqrt{2}\sum_{(k,m)} b_{k,m} \,m\,\Delta_{k,m} \,(r\sin\theta\,\,\Omega^{(1)}\, \sin\hat{v}_{k,m} + \Omega^{(2)}\, \cos\hat{v}_{k,m})\,,
\end{equation}
where
\begin{equation}
\sum_{(k,m)}\equiv \sum_{k=1}^\infty \sum_{m=0}^k
\end{equation}
and
\begin{equation}
\label{modelattice}
\hat{v}_{k,m} \equiv m \frac{\sqrt{2}\,v}{R}+(k-m)\,\phi - m\,\psi + \eta_{km}\,.
\end{equation}
Compared to~\eqref{Z4modes}, we have added a mode-dependent constant phase-shift in the definition of $\hat{v}_{k,m}$, so that \eqref{Z4full} can be thought of as the general Fourier expansion of $Z_4$. We can think of $b_{k,m}$ as the Fourier coefficients of a function of two variables since these modes are related to the $\phi$ and $\psi$ coordinates.

Similarly, for the other pairs $(Z_1, \Theta_2)$ and $(Z_2, \Theta_1)$,
a general class of solutions is given by
\begin{subequations} \label{ZiThetajfull}
 \begin{align} 
 Z_I  &~=~ R\,{b^I_0\over \Sigma}+
 R\,\sum_{(k,m)}b^I_{k,m}\, \frac{\Delta_{k,m}}{\Sigma} \,\cos\hat{v}_{k,m} \,, \label{Zifull} \\
 \Theta_J  &~=~  -\sqrt{2}\sum_{(k,m)} b^I_{k,m} \,m\,\Delta_{k,m} \,(r\sin\theta\,\,\Omega^{(1)}\, \sin\hat{v}_{k,m} + \Omega^{(2)}\, \cos\hat{v}_{k,m})\,\label{Thetajfull}
 \end{align}
\end{subequations}
(for $\{I,J\} = \{1,2\}$), where $b^I_0$, $b^I_{k,m}$ $(I=1,2)$ are new sets of
arbitrary Fourier coefficients. We could also introduce new, independent phase constants, $\eta_{km}^I$,  in (\ref{ZiThetajfull}).  We have thus found a quite general class of solutions to the
first layer of BPS equations (\ref{eqZ1Theta2})--(\ref{eqZ4Theta4}), that can be parameterized by several arbitrary functions of two variables.

As far as the first layer of equations go, the functions
\eqref{Z4full}--\eqref{ZiThetajfull} are  solutions, 
however, it still remains to solve the second layer of equations and
impose regularity on the full geometry.  We will discuss this in detail
in examples in Section \ref{Sect:examples}, but we will not tackle this problem in full generality in this paper.
Our goal here is  to show that there are microstate geometries that
fluctuate as a generic function of two variables.  To that end, we will
simplify the problem by using further insights from the 
rigidly-generated solution discussed in section \ref{ss:particularSoln} and constraining
the form of the Fourier expansions in \eqref{ZiThetajfull}, to obtain a relatively simple family of superstrata
solutions.

\subsection{A three-charge ansatz}
\label{ss:inspiredsol}

In this paper we will make an ansatz in which the Fourier expansions for $(Z_1,\Theta_2)$ and
$(Z_2,\Theta_1)$ are determined in terms of the Fourier expansion of 
$(Z_4,\Theta_4)$.  Because the $b_{k,m}$ will  remain arbitrary, this will still represent a 
solution that depends on a function of two variables.  For simplicity, we will set all the phase constants to
zero:  $\eta_{k,m}=\eta_{k,m}^I=0$.

Our ansatz is inspired by the rigidly-generated three-charge
solution in subsection \ref{ss:particularSoln}.  First, one finds that
this  rigidly-generated solution actually leaves $Z_2$ and $\Theta_1=0$ unchanged
from the two-charge solution.  Thus, we also assume that $Z_2$ is not
deformed and remains as it is in the two-charge solution
(\ref{supertubeplus}). Then, (\ref{eqZ2Theta1}) implies
$\Theta_1=0$. So, we set
 \begin{equation}\label{Z2full}
Z_2= \frac{Q_5}{\Sigma}\,,\quad \Theta_1=0\,.
\end{equation}
Namely, we set $b^{I=2}_{k,m}=0$ for all $k,m$.

 Again, drawing inspiration from the two-charge seed solution (\ref{supertubeplus}), one would expect $Z_1$ to have $v$-dependent terms that are quadratic in $b_{k,m}$ (namely, $b^{I=1}_{k,m}$ will be quadratic in $b_{k,m}$).  A first guess (which will be further substantiated by our analysis in Section~\ref{Sect:layer2}) would be to adjust these terms in such a way that $Z_1 Z_2-Z_4^2$ be non-oscillating.  However, one can immediately see that when $Z_4$ contains more than one mode this is not possible; the product of $Z_4^{(k_1,m_1)}$ and $Z_4^{(k_2,m_2)}$ has the form
 \begin{equation}
\notag 
\Delta_{k_1+k_2,m_1+m_2} \,\bigl(\cos \hat{v}_{k_1+k_2,m_1+m_2} +  \cos\hat{v}_{k_1-k_2,m_1-m_2}\bigr)\,.
\end{equation}
The first term is precisely of the form of the terms that can appear in the mode expansion (\ref{ZiThetajfull}) of $(Z_1,  \Theta_2)$, but the second term is not of this form.  In $Z_1 Z_2-Z_4^2$, it is thus  possible to cancel all the terms proportional to the mode $\hat{v}_{k_1+k_2,m_1+m_2}$, but the modes $\hat{v}_{k_1-k_2,m_1-m_2}$ will remain. As we will see below, arranging this partial cancellation appears to be an important part of regularity of the solution.
These observations motivate the following ansatz:
 \begin{equation}\label{Z1ansatz}
 \begin{aligned}
Z_1&= \frac{R^2}{2\,Q_5}\Biggl[\frac{2 a^2+b^2}{\Sigma}
  +\sum_{(k_1,m_1)} \sum_{(k_2,m_2)} b_{k_1,m_1} b_{k_2,m_2}\,\biggl(\frac{\Delta_{k_1+k_2,m_1+m_2}}{\Sigma}\,\cos\hat{v}_{k_1+k_2,m_1+m_2}\\
&\hspace{35ex}
+c_{k_1,m_1;k_2,m_2}\,\frac{\Delta_{k_1-k_2,m_1-m_2}}{\Sigma}\,\cos\hat{v}_{k_1-k_2,m_1-m_2}\biggr)\Biggr]\,,
\end{aligned}
\end{equation}
where $c_{k_1,m_1;k_2,m_2}$ are coefficients that we will fix by requiring regularity. 
The ansatz for the 2-form, $\Theta_2$, corresponding to this form of $Z_1$ is precisely the appropriate
parallel of (\ref{Theta4full}):
 \begin{align}
\Theta_2 &=  -\frac{R}{\sqrt{2}\,Q_5}
   \sum_{(k_1,m_1)} \sum_{(k_2,m_2)} b_{k_1,m_1} b_{k_2,m_2}\notag\\
 & \times\biggl[(m_1+m_2) \Delta_{k_1+k_2,m_1+m_2}  
 \bigl(r\sin\theta\,\,\Omega^{(1)} \sin\hat{v}_{k_1+k_2,m_1+m_2}
+ \Omega^{(2)} \cos\hat{v}_{k_1+k_2,m_1+m_2}\bigr) \label{Theta2full}
  \\
&\quad+ c_{k_1,m_1;k_2,m_2} (m_1-m_2) \Delta_{k_1-k_2,m_1-m_2}  \bigl(r\sin\theta\,\,\Omega^{(1)} \sin\hat{v}_{k_1-k_2,m_1-m_2} + \Omega^{(2)} \cos\hat{v}_{k_1-k_2,m_1-m_2}\bigr)\biggr]\,,\notag
\end{align}
which  indeed satisfies
(\ref{eqZ1Theta2}).
We assume that the coefficients $c_{k_1,m_1;k_2,m_2}$ are non-vanishing only when the mode $(k_1-k_2,m_1-m_2)$ is allowed: for this one needs $k_1-k_2\not=0$; if we assume, without loss of generality, that $k_1-k_2>0$,  one also needs $k_1-k_2\ge m_1-m_2\ge 0$.  As we will see below, the value of the $c_{k_1,m_1;k_2,m_2}$ will be determined in such a way that the angular-momentum one-form, $\omega$, is regular at the center of the $\mathbb{R}^4$ base space of the solution. The parameter $b$ appearing in the non-oscillating part of $Z_1$ has its origins in the terms with $k_1= k_2$ and $m_1 = m_2$, and  $b$ will be fixed by the regularity of the metric at the supertube position $\Sigma=0$.

\section{The second layer} 
\label{Sect:layer2}

To completely specify the ten-dimensional geometry one must first solve
the second layer of the equations, (\ref{eqomega}) and (\ref{eqcalF}),
and thereby obtain expressions for the one-form, $\omega$, which encodes
the angular momentum, and the function, $\mathcal{F}$, associated with
the momentum charge.  Having done this, one must also impose whatever
constraints are necessary to achieve regularity.

One of our early concerns was that, given our assumptions about the
$v$-independence of the base metric and the one-form, $\beta$, the
regularity constraints might show that there are no generic superstrata
in this class. However, the solution-generating techniques show that
there must be at least a family of non-trivial solutions that are
obtained from rotations of generic shape modes of the D1-D5
configurations.  Such a family would still only be parameterized by
functions of one variable, but our approach is more general: we have
used the solution-generating techniques to find modes that solve the
first layer of equations and we now take arbitrary linear superpositions
of them to generate new families of solutions.

In this and the next sections, we will demonstrate that this approach 
indeed leads to a (smooth) superstratum that fluctuates as a generic
function of two variables.  As will become evident, this is technically
the hardest part of the construction and so we will try to break the
problem into manageable pieces before going into generalities.  In this
section we outline the general structure of the equations that $\omega$
and $\mathcal{F}$ must satisfy, and then, in the subsequent section, we
will give explicit examples illustrating the cancellation of
singularities to demonstrate the existence of families of smooth
solutions.  

Here we concentrate on the regularity constraints that come from the behavior of the metric at the center of $\mathbb{R}^4$, which in our coordinates is at $(r=0,\,\theta=0)$. At this point the angular coordinates $\theta,\phi,\psi$ degenerate, and if a tensor depends on these coordinates and/or has legs along these angular directions, it might be singular even without exhibiting an explicit divergence. The conditions for regularity are analogous to the ones at the center of the plane in polar coordinates. Another possible source of singularities are the terms diverging at the supertube location $(r=0,\,\theta=\pi/2)$. The singularity analysis at this location parallels the one of two-charge solutions and we leave it to Section \ref{Sect:physics}.

\subsection{The system of equations for $\omega$ and $\cF$}
\label{ss:omFeqns}

We  begin with the general mode expansions \eqref{ZiThetajfull} where $(Z_1, \Theta_2)$, $(Z_2,\Theta_1)$  and gradually proceed to our specific ansatz \eqref{Z2full}--\eqref{Theta2full} in which the Fourier coefficients in $(Z_1, \Theta_2)$, $(Z_2,\Theta_1)$ have restricted forms.

Equations (\ref{eqomega}) and (\ref{eqcalF}) form a linear system of
differential equations for $\omega$ and $\mathcal{F}$, and the source
term on the right hand side is a quadratic combination of $Z_I$ and
$\Theta_I$ where $I=1,2,4$.  In  general,   each of $Z_I$ and $\Theta_I$ is a sum over modes
labeled by $(k,m)$ and so the source term will be  a product of two
modes. Linearity means that one can solve these equations
independently for each such pair of modes. We will denote the
contribution to $\omega$ and $\mathcal{F}$ coming from the product of
two modes $(k_1,m_1)$ and $(k_2,m_2)$ by $\omega_{k_1,m_1;k_2,m_2}$ and
$\mathcal{F}_{k_1,m_1;k_2,m_2}$. Thus $\omega$ and $\mathcal{F}$ have
the following general form:
\begin{equation}\label{omegaTOT}
\omega=\omega_0 + \sum_{(k_1,m_1)} \sum_{(k_2,m_2)} \omega_{k_1,m_1;k_2,m_2}\,,~~~~
\mathcal{F}=  \sum_{(k_1,m_1)} \sum_{(k_2,m_2)} \mathcal{F}_{k_1,m_1;k_2,m_2}\,,
\end{equation} 
where $\omega_0$ is the contribution of the round supertube. The product formula of trigonometric functions means that the $v$-, $\psi$- and $\phi$-dependence of $\omega_{k_1,m_1;k_2,m_2}$ and $\mathcal{F}_{k_1,m_1;k_2,m_2}$ will either involve the sum or the difference of the source phases:  $\hat{v}_{k_1+k_2,m_1+m_2}$ or  $\hat{v}_{k_1-k_2,m_1-m_2}$.  Again, linearity means that we may address such pieces separately, so let us analyze the solution of~(\ref{eqomega}) and~(\ref{eqcalF}) for an arbitrary mode whose phase is $\hat{v}_{p,q}$. The form of terms appearing as sources  shows that the full $\omega$ is a linear combinations of contributions of the form 
\begin{align}
\omega_{p,q} & = (\omega_r\,dr+ \omega_\theta\,d\theta)\sin\hat{v}_{p,q} + (\omega_\phi\,d\phi+ \omega_\psi\,d\psi)\cos\hat{v}_{p,q}\,, 
\\
{\cal F}_{p,q}  & =  \frac{2\sqrt{2}}{R}\,W \cos\hat{v}_{p,q}\,,
\end{align}
where $W$ and $\omega_i$, with $i=r,\theta,\phi,\psi$, are functions only
of $r$ and $\theta$. On this ansatz the differential operator that
appears in~(\ref{eqomega}) acts as
\begin{equation}
\mathcal{D}\omega_{p,q} +*_4 \mathcal{D}\omega_{p,q} +\mathcal{F}_{p,q}\,d\beta \equiv \sin\hat{v}_{p,q}\,\Omega^{(1)}\,\mathcal{L}^{(p,q)}_1 + \cos\hat{v}_{p,q}\,(\Omega^{(2)}\,\mathcal{L}^{(p,q)}_2 +\Omega^{(3)}\,\mathcal{L}^{(p,q)}_3 )\,,
\end{equation}
where
\begin{equation}
\begin{aligned}
\mathcal{L}^{(p,q)}_1 &= (r^2+a^2)\cos\theta\,(\partial_r\omega_\theta-\partial_\theta\omega_r)-\frac{q\,r}{\sin\theta}\,\omega_\phi+\frac{1}{r\,\sin\theta}\,(q\,(r^2+a^2)-p\,\Sigma)\,\omega_\psi\,,\\
\mathcal{L}^{(p,q)}_2 &=\frac{r^2+a^2}{r}\,\partial_r\omega_\psi+\cot\theta\,\partial_\theta\omega_\phi + \frac{q\,r\,(r^2+a^2)}{\Sigma}\,\omega_r  
\\ & \qquad+\cot\theta\,\Bigl(\frac{q\,(r^2+a^2)}{\Sigma}-p\Bigr)\,\omega_\theta + 4 a^2 \cos^2\theta \frac{r^2+a^2}{\Sigma^2} W\,,\\
\mathcal{L}^{(p,q)}_3 &=r\,\partial_r\omega_\phi-\tan\theta\,\partial_\theta\omega_\psi +r\,\Bigl(\frac{q\,(r^2+a^2)}{\Sigma}-p\Bigr)\,\omega_r-\frac{q\,r^2\,\tan\theta}{\Sigma}\,\omega_\theta - 4 a^2 \sin^2\theta \frac{r^2}{\Sigma^2} W \,.
\end{aligned}
\end{equation}
The operator in~(\ref{eqcalF}) reduces to\footnote{The unhatted letters
$\mathcal{L}^{(p,q)}_i$ represent scalar quantities while the hatted
$\widehat{\mathcal{L}}^{(p,q)}$ is an operator.}
\begin{equation}
  \label{eq:sDsvd}
  *_4 \mathcal{D}*_4 \Bigl(\partial_v \omega_{p,q}-\frac{1}{2}\mathcal{F}_{p,q}\Bigr) \equiv \frac{\sqrt{2}}{R}\,\cos\hat{v}_{p,q}\,(q\,\mathcal{L}^{(p,q)}_0 + \widehat \cL^{(p,q)} \, W)\,,
\end{equation}
where
\begin{equation}
\begin{aligned}
  \mathcal{L}^{(p,q)}_0 &=-\frac{1}{r\,\Sigma}\,\partial_r (r \,(r^2+a^2)\, \omega_r) -\frac{1}{\Sigma\,\sin\theta\cos\theta}\,\partial_\theta(\sin\theta\cos\theta\,\omega_\theta)\\
&\quad+\frac{1}{\sin^2\theta}\,\Bigl(\frac{p}{r^2+a^2}-\frac{q}{\Sigma}\Bigr)\,\omega_\phi-\frac{q}{\Sigma\,\cos^2\theta}\,\omega_\psi\,,
\end{aligned}
\end{equation}
and the action of the operator $\widehat \cL^{(p,q)}$ on an arbitrary function, $F(r,\theta)$, is defined by:
\begin{align}
\widehat \cL^{(p,q)} \, F   ~&\equiv~  \frac{1}{r\Sigma}\, \partial_r \big( r (r^2 + a^2) \, \partial_r F  \big)  +    \frac{1}{\Sigma\sin \theta \cos \theta}\partial_\theta \big( \sin \theta \cos \theta\, \partial_\theta F  \big)  \nonumber \\
& \hspace{30ex}
- \frac{1}{\sin^2 \theta\Sigma} \, \Big( \frac{p^2\, \Sigma}{r^2 + a^2 } -2\,p q  +  \frac{q^2}{ \cos^2 \theta} \Big)\, F  \label{LapOp1} \\
~&=~    \frac{1}{r\Sigma}\,\partial_r \big( r (r^2 + a^2) \, \partial_r F  \big)  +  \frac{1}{\Sigma\sin \theta \cos \theta}\partial_\theta \big( \sin \theta \cos \theta\, \partial_\theta F  \big)  \nonumber \\
&  \hspace{30ex}
+  \frac{1}{\Sigma} \Big( \frac{p^2 \, a^2}{r^2 + a^2}  -  \frac{(p-q)^2}{\sin^2 \theta}   -  \frac{q ^2}{\cos^2 \theta}    \Big)\, F \,.\label{LapOp2}
\end{align}
Note that $\widehat{\mathcal{L}}^{(0,0)}\equiv \widehat{\mathcal{L}}$ is the scalar Laplacian in the metric~\eqref{ds4flat}. The second expression in~\eqref{LapOp2} shows that this operator is separable.

By using the gauge freedom in~\eqref{gaugecalF} we can set all $v$-dependent modes of ${\cal F}$ to zero and thus we can set $W$ to zero when $q\not=0$.

In terms of the operators defined above, one can show that the parts in
$\omega_{k_1,m_1;k_2,m_2}$ and $\cF_{k_1,m_1;k_2,m_2}$ that depend on
phases $\hat{v}_{k_1\pm k_2,m_1\pm m_2}$ satisfy differential equations
which can be written as the following system of equations:
\begin{equation}
\begin{aligned}\label{eq:system}
&q\,\mathcal{L}^{(p,q)}_0 +  \widehat \cL^{(p,q)} \, W= \frac{R}{\sqrt{2}}\,\frac{\Delta_{k,m}}{\Sigma}\Bigl(\frac{q^2}{\Sigma}+\frac{m^2-q^2}{2\,(r^2+a^2)\,\cos^2\theta}\Bigr)\,,\\
& \mathcal{L}^{(p,q)}_1 =  \frac{R\,q}{\sqrt{2}}\,\frac{r\,\sin\theta\,\Delta_{k,m}}{\Sigma}\,,\qquad
\mathcal{L}^{(p,q)}_2 =   \frac{R\,m}{\sqrt{2}}\,\frac{\Delta_{k,m}}{\Sigma}\,,\qquad \mathcal{L}^{(p,q)}_3 =0\,.
\end{aligned}
\end{equation}
Here $p$, $q$, $k$ and $m$ are integers that depend on the particular
source term in question, and their specific values will be given
below.  The overall coefficients of the right-hand side of
\eqref{eq:system} depend on the particular normalization we choose for
$Z_I$, $\Theta_I$, and they have been chosen for later convenience as we
will explain below.

On the other hand, in our specific ansatz for $(Z_1,\Theta_2)$ and
$(Z_2,\Theta_1)$ given in \eqref{Z2full}--\eqref{Theta2full}, not all of
$Z_I,\Theta_J$ are given by a single sum over modes labeled by $(k,m)$;
some of them contain double sums and some of them contain no sum.
However, by construction, it is still true that the source term
appearing on the right-hand side of Eqs.~\eqref{eqomega} and
\eqref{eqcalF} is a quadratic combination of the coefficients $b_{k,m}$.
Therefore, even for this ansatz, we can solve the equations
independently for each pair of modes, using the mode expansion
\eqref{omegaTOT}.  The resulting equations for a pair of modes again
turn out to be given by the same system of equations \eqref{eq:system},
although the values of $p,q,k,m$  will depend upon the particular source term.  These parameters will  be
are given below.  The overall coefficients of the source on the
right-hand side of \eqref{eq:system} has been conveniently chosen to
correspond to the normalization of $Z_I,\Theta_I$ given in
\eqref{Z2full}--\eqref{Theta2full}.

To summarize, both for the general moding \eqref{ZiThetajfull} and for the
specific ansatz \eqref{Z2full}--\eqref{Theta2full}, the equations for
$\omega$ and $\cF$ can be solved independently for each pair of modes
$(k_1,m_1),(k_2,m_2)$.  Each such pair includes pieces that depend
on different phases $\hat{v}_{p,q}$, and each piece satisfies the system
of equations \eqref{eq:system} with specific values of $p,q,k,m$.
In the next subsection, we analyze the various possibilities that can occur separately,
giving explicit values of the numbers $p,q,k,m$.
For convenience, we define
\begin{align}\label{kpm}
 k_\pm \equiv k_1\pm k_2,\qquad m_\pm \equiv m_1 \pm m_2.
\end{align}

\subsection{The first type of source}
\label{ss:solving1}

For the general mode expansion \eqref{ZiThetajfull}, the fields
$\omega_{k_1,m_1;k_2,m_2}$ and $\mathcal{F}_{k_1,m_1;k_2,m_2}$ contain
terms that depend upon the phase $\hat{v}_{k_+,m_+}$ as discussed above. The
system of equations \eqref{eq:system} for these terms corresponds to the
following values:
\begin{equation}
(p,q)=(k,m)=(k_+,m_+).
\label{system1}
\end{equation}
Remarkably enough, it is easy to guess a solution to the system for
these values of parameters. One can readily
verify that the following is a solution:
\begin{equation}
\omega_{p,q} = \frac{R}{2\sqrt{2}}\,\Delta_{p,q}\,\Bigl(-\frac{dr}{r(r^2+a^2)}\,\sin\hat{v}_{p,q} + \frac{\sin^2\theta\,d\phi+\cos^2\theta\,d\psi}{\Sigma}\,\cos\hat{v}_{p,q}\Bigr)
 \equiv \omega_{p,q}^{(1)}\,.
\label{partsol1}
\end{equation}
Note that the $dr$ part is singular at $r=0$. One might be tempted to
try to remove this singularity by adding a homogeneous solution, but we
have been unable to find one that achieves this. In fact, we believe that there is no regular choice for $\omega_{p,q}$ and, in physically allowed solutions, either  
\begin{itemize}
 \item[(A)] This class of source does not occur, or
 \item[(B)] The singularity must be  canceled by other terms in the full $\omega$.
\end{itemize}

So, in general, we must choose between these two options in order to construct
physically allowed solutions.
Actually, in our specific ansatz \eqref{Z2full}--\eqref{Theta2full} we
have already chosen option (A) to remove the singularity.  To see this recall the mode coefficients $b^{I}_{k,m}$ in $(Z_1,\Theta_2)$
are given by quadratics of the mode coefficients $b_{k,m}$ of $Z_4$,
while $Z_2$ has been kept independent of these modes. This was done so
as to cancel the terms that depended on $\hat{v}_{k_+,m_+}$ in the warp
factor $Z_1 Z_2 - Z_4^2$. One can easily see that this ansatz also means
that the source contributions depending on $\hat{v}_{k_+,m_+}$ in
(\ref{eqomega}) and (\ref{eqcalF}) precisely cancel between terms
quadratic in $(Z_4,\Theta_4)$ and terms linear in $(Z_1,\Theta_2)$.
Namely, in our ansatz, there is no singularity with parameters
\eqref{system1} because the dangerous source terms depending on
$\hat{v}_{k_+,m_+}$ have been arranged to cancel among themselves --- this is what we meant by option (A).

Recall, however, that we have also put in an extra structure in the
ansatz \eqref{Z2full}--\eqref{Theta2full} as terms proportional to
$c_{k_1,m_1;k_2,m_2}$.  They lead to source terms depending on
$\hat{v}_{k_-,m_-}$, which in turn generate contributions to $\omega$
depending on $\hat{v}_{k_-,m_-}$.  This part of $\omega$ is the solution of the
system \eqref{eq:system} with the parameters
\begin{equation}
(p,q)=(k,m)=(k_-,m_-).
\label{system1'}
\end{equation}
For these values of the parameters, the solution is given by
$\omega_{k_-,m_-}^{(1)}$ in \eqref{partsol1} and is singular.  As
discussed at the end of subsection~\ref{ss:inspiredsol}, these singularities
are useful to cancel other singularities arising from other
contributions to $\omega$ discussed below.  Namely, we will choose
option (B) for the source term proportional to $c_{k_1,m_1;k_2,m_2}$.

\subsection{The second type of source}
\label{ss:solving2}

We now restrict to the ansatz  \eqref{Z2full}--\eqref{Theta2full}  and study the remaining
terms in $\omega$ that are dependent upon $\hat v_{k_1-k_2,m_1-m_2}$ and
independent of $c_{k_1,m_1;k_2,m_2}$.  The relevant equations are again
the system \eqref{eq:system}, now with
\begin{equation}
(p,q)=(k_-,m_-),\qquad (k,m)=(k_+,m_+)\,.
\label{system2}
\end{equation}
We will denote this class of solutions by $\omega^{(2)}_{k_-,m_-}$.  The source terms are more complicated to analyze and, while we have succeeded in doing this iteratively, we  have not been able to come up with the general solution. There is, however, one major simplification that we can explicitly use to leverage the rest of the solution in many examples. 
If $q\ne0$, one can use the equations for $\mathcal{L}^{(p,q)}_2$ and $\mathcal{L}^{(p,q)}_3$ to solve algebraically for $\omega^{(2)}_r$ and $\omega^{(2)}_\theta$. One can then eliminate these functions from the other equations and show that 
\begin{equation}
\hat\mu^{(2)}\equiv \frac{1}{2} (\omega_\phi^{(2)} + \omega_\psi^{(2)}) + \frac{R}{4\sqrt{2}}\,\frac{\Delta_{k,m}}{\Sigma}
\end{equation}
satisfies a Poisson equation for the operator $\widehat{\mathcal{L}}^{(p,q)}$ with the choice~\eqref{system2}:
\begin{equation}
\widehat{\mathcal{L}}^{(p,q)}\hat\mu^{(2)} =-\frac{R}{4\sqrt{2}\,q}\,\frac{q\,[(p-q)^2-(k-m)^2]\,\Delta_{k,m+2} +(p-q)(m^2-q^2)\,\Delta_{k,m}}{(r^2+a^2)\,\cos^2\theta\,\Sigma}\,.
\end{equation}
It is convenient to introduce the functions $F_{k,m}^{(p,q)}$ satisfying
\begin{equation}\label{eq:poisson}
 \widehat{\mathcal{L}}^{(p,q)} F_{k,m}^{(p,q)} = \frac{1}{r^2+a^2}\frac{\Delta_{k,m}}{\cos^2\theta\,\Sigma}\,.
\end{equation}
Then one finds
\begin{equation}
\omega_\phi^{(2)} + \omega_\psi^{(2)}=- \frac{R}{2\sqrt{2}}\,\Bigl(\frac{\Delta_{k,m}}{\Sigma} +[(p-q)^2-(k-m)^2]\,F_{k,m+2}^{(p,q)} + \frac{(p-q)(m^2-q^2)}{q}\,F_{k,m}^{(p,q)}\Bigr)\,.
\end{equation}
The recursion relation described in Appendix~\ref{Appendix:recursion}
allows to write $F_{k,m}^{(p,q)}$ explicitly:
\begin{equation}\label{explFkmpq}
F_{k,m}^{(p,q)}=-\frac{1}{4 \,k_1 k_2\,(r^2+a^2)}\,\sum_{s=0}^{k_2-1}\sum_{t=0}^s \binom{s}{t} \frac{\binom{k_1-s-1}{m_1-t-1} \binom{k_2-s-1}{m_2-t-1}}{\binom{k_1-1}{m_1-1} \binom{k_2-1}{m_2-1}}\Delta_{k-2s-2,m-2t-2}\,,
\end{equation}
where we are assuming that $k_1\ge k_2$. 

Thus we \emph{do} have the general solution for one of the components of $\omega$. Moreover, since $\omega^{(2)}_r$ and $\omega^{(2)}_\theta$ are given algebraically in terms of $\omega^{(2)}_\phi$ and $\omega^{(2)}_\psi$, we need just to solve another equation to complete the analysis of~\eqref{eq:system} for this second type of source. We have not been able to simplify this last step in general, but we have been able to solve this equation for several infinite families of solutions and the solutions are series somewhat akin to  (\ref{explFkmpq}).  We therefore expect that there is a general systematic procedure but, as yet, we have not managed to bring it out.  

\subsection{The full $\omega$ and ${\cal F}$}
\label{ss:ocalF}

In the next section we will provide several examples of different non-trivial families of solutions that we hope will clarify the general features of the system of equations~\eqref{eq:system}. From these examples we extract the following general solution-hunting pattern:
\begin{itemize}
\item For all values of $k_i$ and $m_i$, the source terms whose phase is $\hat{v}_{k_+,m_+}$ vanish, as explained in subsection \ref{ss:solving1}.  However, the source terms with phase $\hat{v}_{k_-,m_-}$ remain in general and we have to consider the solutions $\omega^{(2)}_{k_-,m_-}$ discussed in subsection~\ref{ss:solving2}.
\item When $k_-\ge m_- \ge 0$ and $k_->0$, the solution $\omega^{(2)}_{k_-,m_-}$  is singular at $r=0$. In this range of parameters, the full $\omega$ contains also an $\omega^{(1)}_{k_-,m_-}$ contribution discussed at the end of subsection~\ref{ss:solving1}, with a singularity of the same type. Thus the singularities of $\omega^{(1)}_{k_-,m_-}$ and $\omega^{(2)}_{k_-,m_-}$ can be canceled by an appropriate choice of the constant $c_{k_1,m_1;k_2,m_2}$, leaving an $\omega$ which is regular at the center of $\mathbb{R}^4$.
\item For all other values of $k_-$ and $m_-$, when the $\omega^{(1)}$ contribution is absent, there exists a solution for $\omega^{(2)}$ which is by itself regular at the center of space ($r=0, \theta =0$).
\end{itemize}
Thus, the solution to the equations is of the form:
\begin{equation}
  \label{eq:og1}
  \omega_{k_1,m_1;k_2,m_2} = b_{k_1,m_1} b_{k_2,m_2} \,(\omega^{(2)}_{k_-,m_-} + c_{k_1,m_1;k_2,m_2} \, \omega^{(1)}_{k_-,m_-})~.
\end{equation}
As for ${\cal F}$, we know that it can be chosen to be $v$-independent
using the gauge freedom \eqref{gaugecalF} and so it gets contribution
only from $q=0$
\begin{equation}
  \label{eq:calF1}
  {\cal F}_{k_1,m_1;k_2,m_2} = b_{k_1,m_1} b_{k_2,m_1} \,{\cal F}_{k_-,0} ~.
\end{equation}
The term proportional to $c_{k_1,m_1;k_2,m_2}$ is present only when $k_-\ge m_- \ge 0$ and $k_->0$.

In conclusion, for all values of the mode numbers $k_i$, $m_i$, there is a regular solution. We will see that the parameters $b_{k,m}$ which specify the amplitudes of the $(k,m)$ modes inside $Z_4$ are unconstrained, while the constant $c_{k_1,m_1;k_2,m_2}$, which appears in $Z_1$, is uniquely fixed by the regularity requirement.

\section{Examples}
\label{Sect:examples}

We will give in this Section explicit expressions for the contributions, $\omega_{k_1,m_1;k_2,m_2} $ and $\mathcal{F}_{k_1,m_1;k_2,m_2}$, to $\omega$ and $\mathcal{F}$ coming from the modes $(k_1,m_1)$ and $(k_2,m_2)$, for some particular values of $(k_i,m_i)$. We will first consider the terms coming from equal modes: $(k_1,m_1)=(k_2,m_2)$. These contributions are independent of $v$, $\psi$ and $\phi$ and hence are the ones that contribute to the global charges of the geometry. We can  construct the explicit solution for any value of $(k_1,m_1)=(k_2,m_2)$ and we will use it in section~\ref{Sect:physics} to compute the angular momenta and the momentum charge of our superstratum. We will then look at ``oscillating'' contributions produced by unequal modes, which depend on $v$ and/or $\phi$ and $\psi$. We do not know the solution for generic values of $k_1, k_2, m_1, m_2$, but we have constructed several two-parameter families of solutions. We will present two of these  families: the first one shows how the various terms in our Ansatz (\ref{Z1ansatz}) for $Z_1$ crucially conspire to give a regular $\omega$. The second family is rather more intricate and should provide a  representative sample of the computation for generic values of $k_i$ and $m_i$.
 
\subsection{Example 1: $(k_1,m_1)=(k_2,m_2)$}
\label{ss:example1}

For brevity, we will rename these contributions as
$\omega_{k_1,m_1;k_1,m_1}\equiv (b_{k_1,m_1})^2\, \omega_{k_1,m_1}$ and
$\mathcal{F}_{k_1,m_1;k_1,m_1} \equiv (b_{k_1,m_1})^2
\mathcal{F}_{k_1,m_1}$. In general $\omega_{k_1,m_1;k_2,m_2} $
and $\mathcal{F}_{k_1,m_1;k_2,m_2}$ can depend on $\hat{v}_{k_1-k_2,m_1-m_2}$ but this vanishes here, and thus the contributions to $\omega$ and
$\mathcal{F}$ from equal modes are independent of $v$, $\phi$ and
$\psi$.

The equations for $\omega_{k_1,m_1}$ and $\mathcal{F}_{k_1,m_1}$ are obtained from (\ref{eq:system}) by setting $(p,q)=(0,0)$ and $(k,m)=(2 k_1, 2 m_1)$, and can be rewritten as
\begin{gather}\label{eqomegakm}
d\,\omega_{k_1,m_1} + *_4 d\,\omega_{k_1,m_1} + \mathcal{F}_{k_1,m_1} \,d\beta= \sqrt{2}\,R\,m_1\frac{\Delta_{2k_1,2m_1}}{\Sigma}\,\Omega^{(2)}\,,
\\
-*_4 d *_4 d\, \mathcal{F}_{k_1,m_1} \equiv  \widehat{\mathcal{L}} \,  \mathcal{F}_{k_1,m_1} =  \frac{(2 m_1)^2}{r^2+a^2}\,\frac{\Delta_{2 k_1,2 m_1}}{\cos^2\theta\,\Sigma}\,.
\end{gather} 

We have seen in Section~\ref{ss:solving2} that the regular solution of the equation for $\mathcal{F}_{k_1,m_1}$ is
\begin{equation}
\mathcal{F}_{k_1,m_1} = (2 m_1)^2\,F^{(0,0)}_{2k_1,2m_1}\,,
\end{equation}
where the function $F^{(0,0)}_{2k_1,2m_1}$ is obtained from (\ref{explFkmpq}) by setting $(p,q)=(0,0)$. 

In equation (\ref{eqomegakm}) 
we can see by inspection that when $(p,q)=(k_-,m_-)=(0,0)$ the $r$ and $\theta$ components of $\omega$ can be set
to zero. One can then write
\begin{equation}
\omega_{k_1,m_1} = \mu_{k_1,m_1} (d\psi+d\phi) + \zeta_{k_1,m_1}(d\psi-d\phi)\,.
\label{omkmparts1}
\end{equation}
Inspired by the results of \cite{Niehoff:2013kia}, we define
\begin{equation}
\hat \mu_{k_1,m_1} \equiv \mu_{k_1,m_1} +\frac{R}{4\sqrt{2}}\frac{r^2+a^2\sin^2\theta}{\Sigma}\mathcal{F}_{k_1,m_1}+\frac{R}{4\sqrt{2}} \,\frac{\Delta_{2k_1,2m_2}}{\Sigma}\,.
\end{equation}
One can show that $\hat \mu_{k_1,m_1}$ satisfies a Poisson equation of the form of equation (\ref{eq:poisson}):
\begin{equation}
\widehat{\mathcal{L}}\, \hat \mu_{k_1,m_1} =  \frac{R\,(k_1-m_1)^2}{\sqrt{2}(r^2+a^2)}\,\frac{\Delta_{2k_1,2m_1+2}}{\cos^2\theta\,\Sigma}\,.
\end{equation} 
Therefore
\begin{equation}
\hat \mu_{k_1,m_1}=\frac{R}{\sqrt{2}}\,(k_1-m_1)^2\,F^{(0,0)}_{2k_1,2m_1+2} + \frac{x_{k_1,m_1}}{\Sigma}\,,
\end{equation}
where the last term is  harmonic  and the constant $x_{k_1,m_1}$ is determined by regularity as follows: At the center of $\mathbb{R}^4$ ($r=0,\,\theta=0$) the angular coordinates $\psi$ and $\phi$ degenerate, and $\mu_{k_1,m_1}$ must vanish for $\omega_{k_1,m_1}$ to be a regular 1-form. This condition determines $x_{k_1,m_1}$:
\begin{equation}
x_{k_1,m_1}=\frac{R}{4\sqrt{2}}\Bigl[\delta_{k_1,m_1}+\binom{k_1}{m_1}^{\!\!-2}\sum_{s=0}^{k_1-1}\binom{s}{s-(k_1-m_1-1)}\Bigr]=\frac{R}{4\sqrt{2}}\,\binom{k_1}{m_1}^{\!\!-1} \,.
\end{equation}
So we find
\begin{equation}
\mu_{k_1,m_1} =\frac{R}{4\sqrt{2}}\Bigl[\frac{1}{\Sigma}\,\binom{k_1}{m_1}^{\!\!-1} -\frac{\Delta_{2k_1,2m_1}}{\Sigma}+(2k_1-2m_1)^2\, F^{(0,0)}_{2k_1,2m_1+2} - (2m_1)^2\, \frac{r^2+a^2\sin^2\theta}{\Sigma} \,F^{(0,0)}_{2k_1,2m_1}\Bigr]\,.
\end{equation}

The remaining component of $\omega_{k_1,m_1}$ in (\ref{omkmparts1}) is  $\zeta_{k_1,m_1}$ and this can
now be found from  (\ref{eqomegakm}), which gives
\begin{equation}
\begin{aligned}
\partial_r \zeta_{k_1,m_1} &= \frac{r^2 \cos2\theta-a^2\sin^2\theta}{r^2+a^2 \sin^2\theta} \partial_r\mu_{k_1,m_1} -\frac{r \sin2\theta}{r^2+a^2\sin^2\theta}\partial_\theta \mu_{k_1,m_1}\\
&\quad+\frac{\sqrt{2}R\,r\,\sin^2\theta}{\Sigma(r^2+a^2\sin^2\theta)}\Bigl(m_1\,\Delta_{2k_1,2m_1}-\frac{a^2 (2r^2+a^2)\cos^2\theta}{\Sigma}\mathcal{F}_{k_1,m_1}\Bigr) \,,\\
\partial_\theta\zeta_{k_1,m_1}&=\frac{r(r^2+a^2) \sin2\theta}{r^2+a^2\sin^2\theta}\partial_r \mu_{k_1,m_1}+ \frac{r^2 \cos2\theta-a^2\sin^2\theta}{r^2+a^2 \sin^2\theta} \partial_\theta\mu_{k_1,m_1}\\
&\quad-\frac{R\,r^2\,\sin2\theta}{\sqrt{2}\Sigma(r^2+a^2\sin^2\theta)}\Bigl(m_1\,\Delta_{2k_1,2m_1}-\frac{a^2 (r^2+a^2)\cos 2\theta}{\Sigma}\mathcal{F}_{k_1,m_1}\Bigr)\,.
\end{aligned}
\end{equation} 
The relations above can be straightforwardly integrated to give
$\zeta_{k_1,m_1}$.  Although we have not found a general simple
expression for $\zeta_{k_1,m_1}$ for general values of $(k_1,m_1)$,
we have verified its regularity
for several values of $k_1$ and $m_1$.

\subsection{Example 2: $(k_2,m_2)=(1,0)$;  $(k_1,m_1)$ arbitrary}
\label{ss:example2}
Consider a pair of modes with $(k_2,m_2)=(1,0)$ and generic $(k_1, m_1)$ (as usual, one needs $k_1\ge 1$ and $k_1\ge m_1$ for the mode to be allowed).   Here we will simply let $\omega$ denote the contribution to the angular momentum vector from this pair of modes.  Our ansatz for $Z_1$ (see (\ref{Z1ansatz})) contains a term that depends on the difference of the modes ($\hat{v}_{k_-,m_-}$) and the coefficient of this term, $c_{k_1,m_1;k_2,m_2}$, will be abbreviated to $c$  here.   The contribution of this term to  $\omega$ and the corresponding contribution from the similar term in $\Theta_2$ will be denoted by $\omega^{(1)}$ and the  remaining part of $\omega$ will be denoted by $\omega^{(2)}$.   We will also choose the gauge in which the  contribution to $\mathcal{F}$ vanishes.
 
Thus
\begin{equation}
\omega= c\,\omega^{(1)}+ \omega^{(2)}\,.
\end{equation}
Both $\omega^{(1)}$ and $\omega^{(2)}$ are only functions of $\hat{v}_{k_-,m_-}$, which in this subsection we abbreviate as $\hat{v}$:
\begin{equation}
\hat{v}\equiv m_1 \frac{\sqrt{2}\,v}{R}+(k_1-1-m_1)\,\phi - m_1\,\psi\,.
\end{equation}
Examining the equations for $\omega^{(1)}$ and $\omega^{(2)}$ derived from (\ref{eqomega}) and (\ref{eqcalF}), one finds the following solutions
\begin{align}
\omega^{(1)}&= \frac{R}{\sqrt{2}}\,\Delta_{k_1-1,m_1}\Bigl[-\frac{dr}{r(r^2+a^2)} \,\sin \hat{v}+\frac{\sin^2\theta d\phi+\cos^2\theta d\psi}{\Sigma}\,\cos \hat{v} \Bigr]\,,
\\
\omega^{(2)} &= -\frac{R}{\sqrt{2}} \frac{\Delta_{k_1-1,m_1}}{r^2+a^2}\Bigl[\Bigl(\frac{m_1-k_1}{k_1}\frac{dr}{r}-\frac{m_1}{k_1}\tan\theta d\theta\Bigr)\sin \hat{v}\notag\\
&\qquad
 +\Bigl(\frac{r^2+a^2}{\Sigma}\sin^2\theta d\phi + \Bigl(\frac{r^2+a^2}{\Sigma}\cos^2\theta -\frac{m_1}{k_1} \Bigr) d\psi\Bigr) \cos \hat{v}\Bigr]\,.
\label{omega2ex}
\end{align}
We note that generically both $\omega^{(1)}$ and $\omega^{(2)}$ are singular at the center of $\mathbb{R}^4$ ($r=0 , \, \theta=0$). One can however cancel this singularity in the full $\omega$ by choosing
\begin{equation}
c = \frac{k_1-m_1}{k_1}\,.
\end{equation}
For this choice one obtains
\begin{equation}\label{omegaex}
\omega =  -\frac{R}{\sqrt{2}} \frac{\Delta_{k_1-1,m_1}}{r^2+a^2}\Bigl[-\frac{m_1}{k_1}\tan\theta\, d\theta \,\sin \hat{v}+\frac{m_1}{k_1}\frac{(r^2+a^2) d\phi -r^2 d\psi}{\Sigma}\sin^2\theta\, \cos \hat{v} \Bigr]\,,
\end{equation}
which is a regular\footnote{Note that for $\omega$ to be regular at $r=0,\, \theta=0$ it is not sufficient that its components do not diverge. The $\phi$ and $\psi$ components of $\omega$ have to vanish at the center of $\mathbb{R}^4$, where the polar coordinates become degenerate. Moreover, since $\omega$ depends non-trivially on $\phi$ and $\psi$ through the combination $\hat{v}$, its angular components have to vanish at least like $\Delta_{k_1-1,m_1}$. One can see that all these conditions are satisfied by the 1-form in (\ref{omegaex}).} 1-form (excluding the usual singularities at the supertube location $\Sigma=0$, which have to be treated separately).

The solution with $k_1=m_1$ is exceptional: for these modes the last term in $Z_1$ is not allowed (because $k_1-k_2=k_1-1<m_1-m_2=m_1$) and hence the contribution $\omega^{(1)}$ is not present a priori. One can see from (\ref{omega2ex}) that when $k_1=m_1$, $\omega^{(2)}$ is regular by itself.  Note that the final result (\ref{omegaex}) applies also when $k_1=m_1$.

This example shows that the form of $Z_1$ chosen in (\ref{Z1ansatz}) is crucial for the smoothness of the full geometry. In particular, the last term in (\ref{Z1ansatz}) has to be included every time it is allowed and its coefficient is uniquely fixed by the regularity of $\omega$.

The next example will show how this structure extends to more generic values of $k_i$, $m_i$.
\subsection{Example 3: $k_1=m_1+1$, $m_2=1$}
\label{ss:example3}
Consider now the contribution produced by two modes with $k_1=m_1+1$, $m_2=1$ and generic values of $k_2$ and $m_1$. This situation is quite generic  because  all the integers $k_i$, $m_i$ can be different and non-vanishing. As in the previous subsection, we will lighten the notation by suppressing all $k_i$ and $m_i$-dependent indices and work in the gauge with $\mathcal{F}=0$. We  split $\omega$ as $\omega= c\,\omega^{(1)}+\omega^{(2)}$, where the term multiplied by $c$ is the one given in (\ref{partsol1}). The non-trivial task is to determine $\omega^{(2)}$ by solving the system of equations (\ref{eq:system}) with $(p,q)=(m_1+1-k_2,m_1-1)$, $(k,m)=(k_2+m_1+1, m_2+1)$ and to show that its potential singularities can be canceled by $\omega^{(1)}$ for some suitable value of the constant $c$. In this subsection the oscillating factors are functions of
\begin{equation}
\hat{v}\equiv (m_1-1) \frac{\sqrt{2}\,v}{R}+(2-k_2)\,\phi - (m_1-1)\,\psi\,.
\end{equation}

The strategy outlined in Section~\ref{ss:solving2}, and some inspired guesses, lead to the following solution for $\omega^{(2)}$:
\begin{align}
\omega^{(2)}_r &= -\frac{R\,r}{\sqrt{2}\,k_2 (m_1^2-1)}\,\frac{m_1 (k_2+m_1+1) \Delta_{k_2+m_1-1,m_1-1}+(k_2+m_1-1) \Delta_{k_2+m_1-3,m_1-1}}{(r^2+a^2)^2}\,,\nonumber\\
\omega^{(2)}_\theta & = \frac{R}{\sqrt{2}\, k_2 (m_1^2-1) a^2 \sin\theta\cos\theta} \,\Bigl[2(m_1-1) \Delta_{k_2+m_1-3,m_1-1}\nonumber\\
 &\quad+(m_1-1)(m_1-2) \Delta_{k_2+m_1-1,m_1-1}+m_1 (k_2-2) \Delta_{k_2+m_1-1,m_1+1}\nonumber\\
 &\quad-m_1(m_1-1) \Delta_{k_2+m_1+1,m_1-1}+(m_1^2(k_2-1)+1) \Delta_{k_2+m_1+1,m_1+1}\Bigr] ,\nonumber\\
\omega^{(2)}_\phi & = -\frac{R}{\sqrt{2}}\frac{\Delta_{k_2+m_1+1,m_1+1}}{\Sigma}\sin^2\theta -\frac{R}{\sqrt{2}\,k_2 (m_1^2-1) a^2}\,\Bigl[2(m_1-1)\Delta_{k_2+m_1-3,m_1-1}\label{complicatedomega}\\
&\quad +(m_1^2-2m_1+k_2-1)\Delta_{k_2+m_1-1,m_1-1}+m_1 (k_2-2)\Delta_{k_2+m_1-1,m_1+1}\nonumber\\
&\quad +m_1(k_2-m_1-1)\Delta_{k_2+m_1+1,m_1-1}+(k_2(m_1^2+m_1-1)-m_1(m_1+1))\Delta_{k_2+m_1+1,m_1+1}\Bigr]\,,\nonumber\\
\omega^{(2)}_\psi & = -\frac{R}{\sqrt{2}}\frac{\Delta_{k_2+m_1+1,m_1+1}}{\Sigma}\cos^2\theta -\frac{R}{\sqrt{2}\,k_2 (m_1^2-1) a^2}\,\Bigl[(k_2-1)(m_1-1)\Delta_{k_2+m_1+1,m_1+3}\nonumber\\
&\quad -2(m_1-1)\Delta_{k_2+m_1-3,m_1-1}-(m_1-1)(m_1-2)\Delta_{k_2+m_1-1,m_1-1}\nonumber\\
&\quad -(m_1-1)(k_2-3)\Delta_{k_2+m_1-1,m_1+1}+m_1(m_1-1)\Delta_{k_2+m_1+1,m_1-1}\nonumber\\
&\quad -(m_1-1)(m_1(k_2-1)+1)\Delta_{k_2+m_1+1,m_1+1}\Bigr]\,.\nonumber
\end{align}
What interests us about this complicated expression  is its regularity property at the center of $\mathbb{R}^4$ ($r=0,\theta=0$). Remembering the form of $\Delta_{k,m}$, we see that the most stringent regularity constraint comes from the terms $\Delta_{k_2+m_1-3,m_1-1}(\sin\theta)^{-1}$ and $\Delta_{k_2+m_1-1,m_1+1} (\sin\theta)^{-1}$ in $\omega^{(2)}_\theta$:  to avoid a singularity at $\theta=0$ one needs $k_2\ge 3$. Note that this is precisely the range of parameters for which the term proportional to $c$ in the $Z_1$ of (\ref{Z1ansatz}) is not allowed (because $k_1-k_2<m_1-m_2$) and hence the $\omega^{(1)}$ contribution to $\omega$ is absent. So when $k_2\ge 3$ the full $\omega$ coincides with $\omega^{(2)}$ and its explicit expression (\ref{complicatedomega}) shows its regularity.

On the other hand the singularities of $\omega^{(2)}$ for $k_2=1,2$ are expected to be canceled by the $\omega^{(1)}$ term, which is allowed for these values of $k_2$. Comparing the form of $\omega^{(1)}$ in (\ref{partsol1}) with the $\omega^{(2)}$ above, we see however that this cancellation of singularities cannot happen directly: $\omega^{(1)}$ has a singular $r$ component and a vanishing $\theta$ component, while $\omega^{(2)}$ has a singular $\theta$ component. There is however a resolution of this conundrum: when $k_2=1,2$ one can add to $\omega^{(2)}$ a solution of the homogeneous equation which shifts the $\omega^{(2)}$ singularity from the $\theta$ to the $r$ component; moreover the singularity of the new $r$ component is precisely of the type that can be canceled by $\omega^{(1)}$.

For $k_2=2$ the appropriate homogeneous solution is
\begin{equation}
(\omega^\mathrm{hom}_r,\omega^\mathrm{hom}_\theta,\omega^\mathrm{hom}_\phi,\omega^\mathrm{hom}_\psi) = \frac{R}{\sqrt{2}\,(m_1+1)\,a^2}\,\Delta_{m_1-1,m_1-1}\Bigl(\frac{a^2}{r(r^2+a^2)},-\frac{1}{\sin\theta\cos\theta},1,-1\Bigr)\,.
\end{equation}
By replacing $\omega^{(2)}\to \omega^{(2)}+\omega^\mathrm{hom}$ one obtains a new solution for $\omega^{(2)}$ with a regular $\theta$ component and a singular $r$ component. The singularity of the $r$ component comes entirely from $\omega^\mathrm{hom}$. We recall that the $\omega^{(1)}$ contribution to $\omega$ is given in (\ref{partsol1}), where for this value of $k_2$ one has $p=q=m_1-1$. Comparing then $\omega^\mathrm{hom}_r$ with $\omega^{(1)}_r$, we see that the full $\omega=c\,\omega^{(1)}+\omega^{(2)}$ is regular if one picks $c=\frac{2}{m_1+1}$.

For $k_1=1$ the situation is a bit more involved, because even the $r$, $\phi$ and $\psi$ components of $\omega^{(2)}$ diverge at $\theta=0$. The appropriate homogeneous solution to add to $\omega^{(2)}$ is now:
\begin{align}
\omega^\mathrm{hom}_r &=\frac{R\,m_1}{\sqrt{2}(m_1^2-1)}\frac{(m_1-2) \Delta_{m_1,m_1-1}+\Delta_{m_1-2,m_1-1}-\Delta_{m_1,m_1+1}}{r (r^2+a^2)}\,,\nonumber\\
\omega^\mathrm{hom}_\theta &=-\frac{R}{\sqrt{2}(m_1^2-1)}\frac{2(m_1-1) \Delta_{m_1-4,m_1-2}+(m_1-1)(m_1-2)\Delta_{m_1-2,m_1-2}-m_1\Delta_{m_1-2,m_1}}{r^2+a^2}\,,\nonumber\\
\omega^\mathrm{hom}_\phi &=\frac{R}{\sqrt{2}(m_1^2-1)}\frac{2(m_1-1) \Delta_{m_1-2,m_1-1}+m_1(m_1-2)\Delta_{m_1,m_1-1}-m_1\Delta_{m_1,m_1+1}}{a^2}\,,\\
\omega^\mathrm{hom}_\psi &=-\frac{R\,(m_1-1)}{\sqrt{2}(m_1^2-1)}\frac{2\Delta_{m_1-2,m_1-1}+(m_1-2)\Delta_{m_1,m_1-1}-2\Delta_{m_1,m_1+1}}{a^2} \nonumber\,.
\end{align}
One can check that the new solution $\omega^{(2)}+\omega^\mathrm{hom}$ is now regular with the exception of the $r$ component, which is given by
\begin{equation}\label{ototr}
\omega^{(2)}_r+\omega^\mathrm{hom}_r= -\frac{R\,m_1}{\sqrt{2}\,(m_1^2-1)}\,\Delta_{m_1,m_1-1}\,\frac{3r^2-(m_1-1)a^2}{r (r^2+a^2)^2}\,.
\end{equation}
Recall that $\omega^{(1)}$  is given by (\ref{partsol1}) with $(p,q)=(m_1,m_1-1)$ and hence its $r$ component has precisely the same form as (\ref{ototr}) in the limit $r\to 0$. One can thus take $c=\frac{2 m_1}{m_1+1}$ and obtain a total $\omega$ free of singularities.

\section{Regularity, asymptotically-flat superstrata and their charges}
\label{Sect:physics}

Up to this point, we have focused on the regularity of the metric at the center of $\mathbb{R}^4$, which in our coordinates is at $r=0$, $\theta=0$. The metric coefficients are also singular at the supertube location, $\Sigma=0$. The resolution of these singularities is familiar from the study of the rigid two-charge supertube geometry: there are potentially singular terms in the ten-dimensional metric proportional to $(d\psi+d\phi)^2$ and the condition that guarantees the cancellation of these singularities is
\begin{equation}
\lim_{\Sigma\to 0} \Sigma\Bigl[-\frac{2\,\alpha}{\sqrt{Z_1 Z_2}}\,\beta_0 \,\Bigl(\mu + \frac{1}{2} \,\mathcal{F}\,\beta_0\Bigr)+ \frac{a^2}{4}\,\sqrt{Z_1 Z_2}\Bigr]=0\,,
\end{equation}
with $\beta_0 \equiv (\beta_\psi+\beta_\phi)/2$ and $\mu\equiv (\omega_\psi+\omega_\phi)/2$, where $\omega$ now stands for the total $\omega$ of (\ref{omegaTOT}). This condition fixes the value of the parameter $b$ which appears in the non-oscillating part of $Z_1$, in terms of the mode amplitudes $b_{k,m}$. One finds
\begin{equation}
\label{supertuberegularity}
b^2 = \frac{4\sqrt{2}}{R}\,\sum_{k,m} b^2_{k,m}\,x_{k,m} = \sum_{k,m} b^2_{k,m} \,\binom{k}{m}^{\!\!-1} \,.
\end{equation}

The family of geometries we have constructed thus far are asymptotic (for large $r$) to AdS$_3\times S^3 \times T^4$. These  solutions  can therefore be identified with microstates of the D1-D5 CFT\@. On the other hand, in order to create a geometry that looks like a five-dimensional black hole one needs to have a geometry whose large-distance asymptotic structure is $\mathbb{R}^{4,1}\times S^1\times T^4$ (we will call such geometries ``asymptotically flat''). If we want to identify our solutions with black-hole microstates, it is necessary to show that they can be extended to such asymptotically flat geometries. This requires re-inserting the ``1'''s in the warp factors $Z_1$ and $Z_2$:
\begin{equation}\label{addone}
Z_1 \to 1+ Z_1  \,,\quad Z_2 \to 1+Z_2 \,.
\end{equation}
Note that $Z_4$ remains unchanged.

This modification of $Z_1$ and $Z_2$ adds new source terms to the BPS system: the $\omega$ equations (\ref{eqomega}), (\ref{eqcalF}) indeed imply that the change (\ref{addone}) necessarily generates a modification of $\omega$ of the form
\begin{equation}
\omega\to \omega + \delta\omega\,,
\end{equation}
where $\delta\omega$ satisfies
\begin{equation}\label{eqdeltaomega}
\mathcal{D} \,\delta\omega + *_4  \mathcal{D}\,\delta\omega = \Theta_2\,,\quad *_4\mathcal{D} *_4 \delta \omega=  \dot{Z}_1 \,.
\end{equation}
As usual one can choose a gauge in which $\mathcal{F}$ is unmodified.  

Finding the general solution to this equation is straightforward: if one takes the general form of the $Z_1$ and $\Theta_2$ from (\ref{ZiThetajfull}) one can solve (\ref{eqdeltaomega}) mode by mode.  If we denote by $\delta\omega_{k,m}$ the contribution to $\delta\omega$ from the $(k,m)$ mode in (\ref{ZiThetajfull}), we find
\begin{equation}\label{partsol0}
\delta\omega_{k,m}=\frac{b^{1}_{k,m}\,m}{\sqrt{2}\,k}\,\Delta_{k,m}\,\Bigl[\Bigl(-\frac{d r}{r}+ \tan\theta\,d\theta\Bigr)\,\sin\hat{v}_{k,m} + d\psi\, \cos\hat{v}_{k,m}\Bigr]\,.
\end{equation}
Note that, once again, there is a singularity at $r=0$. The removal of this singularity can be done following a systematic procedure: One must first collect all the terms that give rise to them via (\ref{partsol0})  and (\ref{partsol1}).  As before this will result in quadratics in the Fourier coefficients, $b_{km}$, appearing in  ($Z_4$, $\Theta_4$) and a (now somewhat modified) linear dependence upon the Fourier coefficients $b^1_{km}$ appearing in ($Z_1$, $\Theta_2$) (see  (\ref{Zifull}) and (\ref{Thetajfull})).  One can then solve for these Fourier coefficients and remove the singularities.   

We are not going to investigate asymptotically flat superstrata any further in this paper because it will take us  into another rather technical discussion.  We note that there are some simple examples of asymptotically-flat superstrata in \cite{Giusto:2013bda}  and we will leave the construction of families of asymptotically-flat superstrata to subsequent work \cite{BGRSW:2015}.   Our primary focus for most of the rest  of this paper will be the  examination of CFT states that are holographically dual to the superstratum excitations and for this we only need the somewhat simpler classes of solutions in the ``decoupling limit,'' where we drop the $1$'s.

For the purpose of computing the asymptotic charges of the solution, all the new  terms arising from (\ref{addone}) and the concomitant cancellation of the  singular modes are irrelevant because they are proportional to non-trivial Fourier modes in $v$ and so vanish when integrated over the $S^1$ compact direction. The angular momentum of the geometry can thus be derived from the ``near-horizon'' $\omega$ computed in the previous sections and, in particular, from the $v$-independent contributions generated by equal modes.

The quantized angular momenta $j$ and $\jt$ are given by
\begin{equation}
j = \frac{V_4\,R}{(2\pi)^4 g_s^2 \alpha'^4}\,J\,,\quad \jt = \frac{V_4\,R}{(2\pi)^4 g_s^2 \alpha'^4}\,\tilde J\,,
\end{equation}
where $V_4$ is the volume of $T^4$, $g_s$ is the string coupling and the dimension-full parameters $J$ and $\Jt$ can be extracted from the large radius behavior of the geometry  as:
\begin{equation}
\frac{\beta_0+\mu}{\sqrt{2}} = \frac{J - \Jt\,\cos2\theta}{2\,r^2} + O(r^{-3})\,.
\end{equation}
For our solution we find
\begin{equation}
J= \frac{R}{2}\,\Bigl[a^2+\sum_{k,m} b_{k,m} ^2\, \frac{m}{k}\,\binom{k}{m}^{\!\!-1}\Bigr]\,,\quad \Jt = \frac{R}{2}\,a^2\,.
\end{equation}
Moreover the D1 supergravity charge can be extracted from the large distance behavior of the warp factor $Z_1$ and is given by 
\begin{equation}\label{eq:q1q5ab}
Q_1 = \frac{R^2}{Q_5} \,\Bigl(a^2+\frac{b^2}{2}\Bigr) =\frac{R^2}{Q_5} \,\Bigl[a^2+\frac{1}{2}\, \sum_{k,m} b^2_{k,m} \,\binom{k}{m}^{\!\!-1}\,\Bigr]  \,.
\end{equation}
The D5 supergravity charge, $Q_5$, is not affected by the superstratum fluctuations we consider. The integer numbers $n_1$ and $n_5$ of D1 and D5 branes are related to $Q_1$ and $Q_5$ by the relation
\eqref{Q1Q5_n1n5}.
Altogether we find that the quantized angular momenta of our solution are
\begin{equation}\label{eq:JtJgr}
j = \frac{{\cal N}}{2}\,\Bigl[a^2 +\sum_{k,m} b_{k,m} ^2\, \frac{m}{k}\,\binom{k}{m}^{\!\!-1}\Bigr]\,,\quad \jt =  \frac{\cal N}{2}\,a^2\,.
\end{equation}
with 
\begin{equation}
{\cal N}\equiv \frac{n_1 n_5\,R^2}{Q_1 Q_5} = \frac{n_1 n_5}{a^2+\frac{1}{2}\, \sum_{k,m} b^2_{k,m} \,\binom{k}{m}^{\!\!-1}}\,.
\end{equation}
A similar computation can be performed to derive the momentum charge of the solution. From the geometry we can extract the supergravity momentum charge $Q_p$ as
\begin{equation}
-\frac{\mathcal{F}}{2} = \frac{Q_p}{r^2} + O(r^{-3})\,.
\end{equation}
Our geometry gives
\begin{equation}
Q_p = \frac{1}{2}\,\sum_{k,m} b_{k,m} ^2\, \frac{m}{k}\,\binom{k}{m}^{\!\!-1}\,,
\end{equation}
and hence its quantized momentum charge is 
\begin{equation}\label{eq:npgr}
n_p = \frac{R^2\,V_4}{(2\pi)^4\,g_s^2\, \alpha'^4}\,Q_p = \frac{\cal N}{2}\,\sum_{k,m} b_{k,m} ^2\, \frac{m}{k}\,\binom{k}{m}^{\!\!-1}\,.
\end{equation}
In particular, we note that from (\ref{eq:JtJgr}) and (\ref{eq:npgr}) we have
\begin{equation}\label{eq:npjj}
 n_p ~=~  j -\tilde j \,.
\end{equation}
In the next section we will use the values of $j$, $\jt$ and $n_p$ to help determine the map between our geometries and the dual CFT states.

\section{The CFT description}
\label{Sect:D1D5CFT}

The geometries constructed in the previous sections are asymptotically
AdS$_3 \times S^3 \times T^4$.  According to the general AdS/CFT
paradigm, we expect that they correspond to semi-classical states in the
dual two-dimensional CFT, commonly called the D1-D5 CFT, with a large
central charge $c=6N$ where 
\begin{equation}
 N\equiv n_1 n_5.
\end{equation}
In this section we briefly
recall the main features of the dual CFT that are relevant here\footnote{For more details of the 
D1-D5 CFT, see e.g.\ \cite{Avery:2010qw}
and references therein.}  and give a general description in the CFT
language of the class of states dual to the superstratum geometries that we have constructed.  

\subsection{Basic features of the dual CFT}

The CFT we are interested in has $\cN=(4,4)$ supersymmetry with the
$\cR$-symmetry group $SO(4)_\cR\cong SU(2)_L \times SU(2)_R$ which, in the gravity dual, is
identified with the rotations of the non-compact $\bbR^4$ coordinates
$x^i$ in the space transverse to all the branes. At a special point in its moduli
space, this CFT can be described by a sigma model whose target space is
the orbifold, $(T^4)^N/S_N$, where $S_N$ is the permutation group on $N$
elements.  Namely, we have $N$ copies of 4 free compact bosons and 4
fermions, identified under permutations of the copies.
The $4N$ bosons are labeled \cite{Avery:2010qw, Giusto:2013bda} as
$X^{\dot{A} A}_{(r)} (z,\bar{z})$, where $r =1, \dots, N$ is the copy
index of the $T^4$ and $A,\dot{A}=1,2$ are spinorial indices for the
$SO(4)_I=SU(2)_1\times SU(2)_2$ of the tangent space of $T^4$. The left-
and right-moving fermions are labeled as $\psi_{(r)}^{\alpha
\dot{A}}(z)$ and $\psit_{(r)}^{\dot{\alpha} \dot{A}}(\bar{z})$ where
$\alpha,\dot{\alpha}=\pm$ are spinorial indices for the $\cR$-symmetry
$SU(2)_L \times SU(2)_R$.  Namely, under the $\cR$-symmetry, the bosons
are singlets, while the left- and right-moving fermions transform as
$({\bf 2},{\bf 1})$ and $({\bf 1},{\bf 2})$, respectively.

\begin{figure}[htbp]
\begin{quote}
 \begin{center}
 \includegraphics[height=4cm]{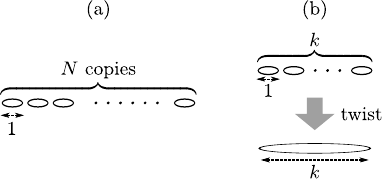}
 \caption{\label{fig:N_copies} \sl 
 (a) The CFT at the orbifold point can be thought of as made of $N$
 copies, each of which contains 4 free bosons and 4 free fermions.  Each
 circle in the figure corresponds to a single copy.
 (b) A twist field intertwines $k$ copies
 into a single strand of length $k$.}
 \end{center}
\end{quote}
\end{figure}

It is useful to visualize the CFT states by representing the $N$ copies, indexed by $(r)$, as $N$
strings  (see  Figure  \ref{fig:N_copies}(a)), on each of which live 4 bosons and 4 fermions:
\begin{equation}
(X^{\dot{A} A}_{(r)}(z,\bar{z})\,, \psi_{(r)}^{\alpha
\dot{A}}(z)\,, \psit_{(r)}^{\dot{\alpha} \dot{A}}(\bar{z})) \,.
\label{Xpsipsit}
\end{equation}
  Besides the operators that can be built
explicitly in terms of the free bosons and fermions, the CFT contains
also \emph{twist} fields that glue together $k$ copies of the free
fields into a single \emph{strand} of length $k$. (See  Figure
\ref{fig:N_copies}(b).)  For this reason, the free point in this CFT
moduli space is usually called the orbifold
point.

On a strand of length $k$, the $k$ copies of the fields are
cyclically glued together and therefore they are $2\pi k$ periodic
instead of $2\pi$ periodic.  This means that the mode numbers of the fields
on a strand of length $k$ are $1/k$ times the mode numbers on a string of length
one.  General states have multiple strands of various lengths.
For instance, if the $\ell^{\rm th}$ strand has length $k$, the mode expansion of 
the fermion  field $\psi_{\ell}(z)$ living on it is
\begin{equation}
  \label{eq:psiell}
  \psi_{\ell}(z) = \sum_{n\in\bbZ} \bigl(\psi_{\frac{n}{k}}\bigr)_{\ell}\, z^{-\frac{n}{k}-\frac{1}{2}}~.
\end{equation}

By construction, the excitations of the bosons, $X^{\dot{A} A}_{(r)}$,
only involve motions in the compactified ($T^4$) directions, whereas the
fermionic excitations carry polarizations ($\cR$-charge) that must be
visible within the six-dimensional space-time.  More concretely, 
the modes of the currents
\begin{align}
J_{\ell}^{\alpha \beta}(z)  & ~\equiv~  \frac{1}{2} \psi_{\ell}^{\alpha \dot{A}}(z)  \, \epsilon_{\dot{A}\dot{B}} \,  \psi_{\ell}^{\beta \dot{B}}(z)  \,,  \qquad \tilde J_{\ell}^{\dot{\alpha} \dot{\beta}}(\bar{z})   ~\equiv~  \frac{1}{2} \tilde\psi_{\ell}^{\dot{\alpha} \dot{A}}(\bar{z})  \, \epsilon_{\dot{A}\dot{B}} \,  \tilde\psi_{\ell}^{\dot{\beta} \dot{B}} (\bar{z}) \,, \label{Jdefn}
\end{align}
can be viewed as bosonizations of the fermions, and because these currents lie entirely in spatial directions of the six dimensional space-time it follows that suitably coherent excitations created by these currents  will  be visible within  six-dimensional supergravity \cite{Bena:2014qxa}.

Note that one should not confuse the labels $(r)$ and $\ell$:  The former labels each  set of bosons and fermions  (\ref{Xpsipsit}) before orbifolding whereas $\ell$ indexes the strands and so labels sets of bosons and fermions that have been orbifolded together to make a  longer effective string. Thus the   currents in (\ref{Jdefn}) are defined for each \emph{individual} strand labeled by $\ell$ and thus give a current algebra of level $k$,  the length of the strand, rather than level $1$, which would be the level of the current algebra for each individual set of fermions in  (\ref{Xpsipsit}).

One can also write the current algebra of the $\cR$-symmetry  by summing over all the fermions or over all the 
 individual currents over all strands:
\begin{align}
J^{\alpha \beta}(z)  ~\equiv~  &  \frac{1}{2} \,\sum_{(r)} \,  \psi_{(r)}^{\alpha \dot{A}}(z)  \, \epsilon_{\dot{A}\dot{B}} \,  \psi_{(r)}^{\beta \dot{B}}(z)   ~=~ \sum_\ell \, J_{\ell}^{\alpha \beta}(z)   \,,  \\
 \tilde J^{\dot{\alpha} \dot{\beta}}(\bar{z})  ~\equiv~  &  \frac{1}{2} \,\sum_{(r)} \,  \tilde\psi_{(r)}^{\dot{\alpha} \dot{A}}(\bar{z})  \, \epsilon_{\dot{A}\dot{B}} \,  \tilde\psi_{(r)}^{\dot{\beta} \dot{B}} (\bar{z}) ~=~
  \sum_{\ell} \, \tilde J_{\ell}^{\dot{\alpha} \dot{\beta}}(\bar{z})   \label{CFTUones}   \,.
\end{align}
This current algebra has level $N$. The standard angular momentum operators, $J^i$ with $i=3,\pm$, are given in
terms of the $J^{\alpha \beta}$ by:
\begin{equation}
\begin{split}
 J^3 &~=~ J^{1 2} ~=~ J^{2 1}    \,,  \qquad 
 J^+  ~=~ J^{1 1}\,,  \qquad 
 J^-  ~=~ J^{2 2}\,,
\end{split}
\end{equation}
and likewise for $\Jt^i$.  Also for the individual currents we similarly
define $J^i_{\ell}$ and $\Jt^i_{\ell}$, from $J^{\alpha\beta}_{\ell}$ and
$\tilde J_{\ell}^{\dot{\alpha} \dot{\beta}}(\bar{z})$.

Even if the free description of the CFT lies outside the regime where
supergravity is a reliable approximation, it is still a very valuable framework for describing the
states dual to the superstratum.  As usual, supersymmetry is responsible for this utility:   the conformal
dimensions of states preserving $1/8$ of the total $32$ supercharges
and their 3-point correlators~\cite{Baggio:2012rr} are protected and
so, for these observables, it makes sense to match directly the CFT
results obtained at the orbifold point to those derived in the
supergravity description.  A detailed comparison between these two
pictures for the class of states described in this paper deserves a
separate paper following the spirit of what was done
in~\cite{Kanitscheider:2006zf,Kanitscheider:2007wq} for the \nBPS{4} 
states. Here we will provide
just the basic features of the duality between CFT states and bulk geometries.

\begin{figure}[htbp]
\begin{quote}
 \begin{center}
 \includegraphics[height=3cm]{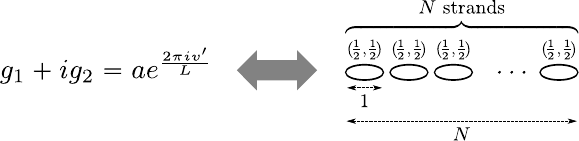}
 \caption{\label{fig:RR_gnd0} \sl 
  The state/geometry dictionary for the maximally-spinning supertube with dipole charge 1.  
  The circular profile given in \eqref{circle} on the gravity side
 (shown on the left) corresponds to the CFT state with $N$ strands all
 with length one and R-charge eigenvalues
 $(j_\ell,\jt_\ell)=(\half,\half)$ (shown on the right).
 }
 \end{center}
\end{quote}
\end{figure}

As one would expect, the first entry in the dictionary maps the global AdS$_3
\times S^3 \times T^4$ solution to the $SL(2,\bbC)$ invariant vacuum in
the NS-NS sector. However, we are interested in states in the
RR sector, which correspond to geometries that can be glued to
an asymptotically $\mathbb{R}^{1,4} \times S^1 \times T^4$ region. The round
supertube solution specified by the profile~\eqref{circle} is the
simplest of such RR states. In order to find the CFT description for
this state, it is sufficient to relate the change of
variables~\eqref{spectralflow} to the spectral flow on the CFT side.   We
first choose a $U(1) \times U(1)$ subgroup of the $\cR$-symmetry group, and
refer to the corresponding currents as $J^3$ and $\Jt^3$ (these
currents correspond to the two $U(1)$ rotation symmetries in the
$\bbR^4$ for the round supertube solution) and their modes as $J^3_n$
and $\Jt^3_n$.  Then, we simply perform a spectral flow of the
NS-NS vacuum state to the RR sector by using $J^3$ and $\Jt^3$. In this
way we obtain an eigenstate of $(J^3_0, \Jt^3_0)$ with eigenvalues
equal to $({N\over 2},{N\over 2})$. At the orbifold point, it is
possible to write the $J^i_0$ and $\Jt^i_0$ as the sum of
generators acting on the $\ell$-th strand, $(J^i_0)_\ell^{}$ and
$(\Jt^i_0)_\ell^{}$.  To avoid clutter,  we define 
$j_\ell\equiv(J^3_0)_\ell^{}$, $\jt_\ell\equiv (\Jt^3_0)_\ell^{}$.
Then, in the free CFT limit, the state is composed of $N$ independent
strands, each one with eigenvalues $(j_\ell,\jt_\ell)=(\half,\half)$. 
This type of strands is annihilated by the modes $(\psi^{+\dot{A}}_0)_\ell^{}$. For a visual explanation of this correspondence, see Figure \ref{fig:RR_gnd0}.

We can now build a dictionary between the supergravity solutions
discussed earlier and a set of pure semi-classical
states in the CFT\@. We  parallel our approach to the supergravity solution
by starting with a review of the  \nBPS{4}  semi-classical states and their descendants. We then 
move to the CFT description of the fluctuating superstrata geometries by examining precisely how
we added momentum modes to the  \nBPS{4}  supertubes.

\subsection{\nBPS{4} states and their descendants}
\label{ss:CFT1mode}

The semi-classical RR ground states that are dual
to \nBPS{4}  geometries were discussed in detail
in~\cite{Kanitscheider:2006zf,Kanitscheider:2007wq} and here we will review the previous results in a language
that is convenient for the generalization in the next section. 

All \nBPS{4}  geometries are determined by a closed profile $g_A(v')$ in
$\mathbb{R}^8$ but, as mentioned above, we focus only on a profile in
an $\mathbb{R}^5$ subspace in order to have states that are invariant under
rotations of the $T^4$ coordinates. Thus, on the geometry side, we have
five periodic functions $g_A(v')$, $A=1,\dots,5$ that can be Fourier
expanded in modes. By using the language of the orbifold free field
description, we can characterize the properties of the profile on the
CFT side as follows: the mode numbers of the Fourier expansion
correspond to the lengths of the strands, the different components
($A=1,\dots,5$) of the profile determine the quantum numbers of each
strand under the $SU(2)_L\times SU(2)_R$ $\cR$-symmetry generators, and
finally the amplitude of each Fourier mode is related to the number of
strands of a particular type present in the dual CFT state. Since we are
focusing on \nBPS{4}  states, each strand has the lowest eigenvalue for
both $(L_0)_\ell^{}$ and $(\tilde{L}_0)_{\ell}$ and so the same is true for
the full state.

\begin{figure}[htbp]
\begin{quote}
 \begin{center}
  \hspace*{-.5cm}
 \includegraphics[height=2.5cm]{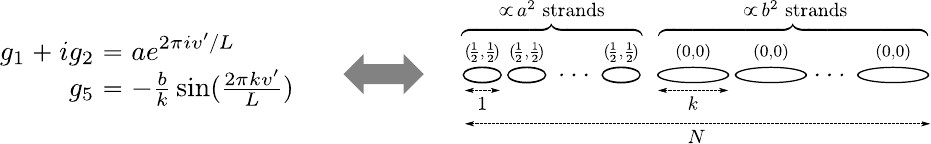}
 \caption{\label{fig:RR_gnd1} \sl 
The state/geometry dictionary for the 1/4-BPS states on which we will add momentum to create superstrata. The profile given in \eqref{circleplus} on the gravity side (shown
 on the left) corresponds to the CFT state with two types of strands
 (shown on the right).  The first type of strand has length one and
 $(j_\ell,\jt_\ell)=(\half,\half)$ while the second type of
 strand has length $k$ and $(j_\ell,\jt_\ell)=(0,0)$.
 }
 \end{center}
\end{quote}
\end{figure}

The profile in  \eqref{circleplus} represents a non-trivial ``deformation'' 
of the simple vacuum state represented by the profile \eqref{circle} whose CFT interpretation, as discussed above,
can be  thought of as $N$ strands of length $1$.  The  profile~\eqref{circleplus} has an extra non-trivial component, $g_5$, that has been added to the 
functions $g_1$ and $g_2$ that are already present in~\eqref{circle}.  It should therefore correspond
to a state with two types of strands:  the standard strands with
$(j_\ell,\jt_\ell)=(\half,\half)$ that are the basic ingredients
 of the state dual to the round supertube, and a second type of
strand  that is obtained from the first by acting with the operator $O_\ell= (\psi^{-\dot{A}})_\ell^{} (\tilde{\psi}^{\dot{-}\dot{B}})_\ell^{} \epsilon_{\dot{A}\dot{B}}$. This operator is a scalar under rotations of the $T^4$ and carries $(j_\ell,\jt_\ell)=(-\half,-\half)$, so that the new strands have quantum numbers $(j_\ell,\jt_\ell)=(0,0)$ and are also invariant under $\mathbb{R}^4$ rotations. Thus it is natural to associate these new strands to the component $g_5$ of the profile. The coefficients $a$ and $b$ determine the
number of the constituent strands of the first and the second type:
$a^2$ is proportional to the number of strands of the first type
and $b^2/(2 k)$ is proportional to the number of strands of the second
type.\footnote{As discussed in~\cite{Kanitscheider:2006zf,
Kanitscheider:2007wq}, this is not the exact characterization of the
dual semi-classical state, even in the large $n_1 n_5$ limit: in general
the dual state is a linear combination of many terms, that is peaked
around the configuration described in the text, with a spread
determined by the coefficients $a$ and $b$. } Note that this is consistent with the relation~\eqref{eq:abq1q5}, since the total length of the state is fixed in terms of $N$. Finally the Fourier mode numbers $k$ of the various components of the profile determine the total length of the corresponding type of strands. We consider states for which the $(\half,\half)$ strands have length $1$, since this is the only Fourier mode present in $g_1$ and $g_2$ and the $(0,0)$ strands have arbitrary length, $k$.\footnote{Note that the geometries dual to these states do not have any conical defects even if the corresponding CFT state has strands of length $k>1$. This is to be contrasted with the
examples considered previously in the literature where all the
components of the profile had the same Fourier mode $k$.}
For a pictorial explanation of 
this correspondence, see Figure \ref{fig:RR_gnd1}.

\begin{figure}[htbp]
\begin{quote}
 \begin{center}
 \includegraphics[height=6cm]{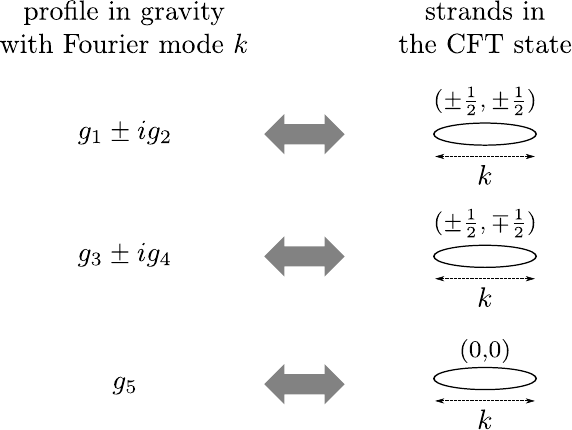}
 \caption{\label{fig:mapping} \sl 
 The state/geometry dictionary for more general 1/4-BPS states.  The Fourier components of
  profiles $g_A$ with mode number $k$ in gravity (shown on the left)
  correspond to the strands in the CFT states with length $k$ with
  specific values of the R-charge $(j_\ell,\jt_\ell)$ (shown on
  the right).  }
 \end{center}
\end{quote}
\end{figure}

In a similar way, it is possible to map different Fourier modes of
each profile components to CFT strands with particular $SU(2)_L\times
SU(2)_R$ quantum numbers. The components $g_1 \pm i g_2$ of the profile
correspond to $(\pm \half,\pm \half)$ strands and the components $g_3 \pm i
g_4$ correspond to $(\pm \half,\mp \half)$ strands. Together with the
correspondence for the component $g_5$ discussed in the example above,
this completes the dictionary between Fourier modes and strand types;
see Figure \ref{fig:mapping} for a visual explanation.
Of course, supergravity solutions correspond to semi-classical states
where each type of strand appears in many copies so as to be suitably
coherent.  The only relevant information for defining the dual state on
the CFT side is the distribution of the numbers of each type of strand in
the full state. Order one variations
from the states discussed above are not visible within the supergravity
limit. 

At this point it is straightforward to extend this correspondence to
descendant states: Both on the bulk and on the CFT side one just
needs to act on the same \nBPS{4}  states with certain generators of the
superconformal algebra.  This programme was initiated
in~\cite{Mathur:2003hj} and a general discussion at the linearized
level can be can be found in~\cite{Mathur:2011gz}. In this paper have we
focused on the $\cR$-symmetry generators.  As summarized in
Section~\ref{ss:particularSoln}, a first example of a
non-linear descendant geometry can be constructed simply by acting
with the exponential $e^{\chi (J^+_{-1}-
  J^-_1)}$~\cite{Giusto:2013rxa,Giusto:2013bda}. Clearly this operation brings about
a non-trivial momentum charge, as the action of each $J^+_{-1}$
obtained by expanding the exponential increases the momentum by one
unit and the average momentum of the descendant state is
determined by the rotation parameter, $\chi$ (See~\cite{Giusto:2013bda}
for the explicit matching of the momentum and the angular momentum
expectation values between the bulk and the CFT descriptions).

It is very important to understand the commonalities and differences between the
construction of the ``rigidly-generated'' states obtained by the rotation above and  our generic 
superstratum fluctuations.   In the orbifold CFT language the ``rigidly-generated'' states  contain 
not only the strands that were present in the original \nBPS{4}  states before the rotation but also
have a new type of momentum-carrying strand that is obtained by acting with the
superalgebra generators involved in the rotation.   The relative number
of the two types of strands ({\it i.e.}\ the RR ground states and the
momentum-carrying ones) is determined by the rotation parameter.  
On the other hand, to make a fluctuating superstratum we rebuilt a complete supergravity solution
starting from almost\footnote{Modulo the constraints imposed by regularity.} arbitrary superpositions
of the linearized forms of all possible ``rigidly-generated'' states and thereby generated far richer
families of CFT states.  As we will see in the next subsection, the rigid rotation is crucial to developing 
the holographic dictionary for each individual mode and in this way we will obtain the CFT dual of 
the generic superstratum geometry.

\subsection{A class of superstrata: the CFT description}
\label{ss:CFTnmode}

We do not, yet, have an exhaustive description of the \nBPS{8} 
geometries as we do for the \nBPS{4}  ones. So it is easier to construct
the dictionary between supergravity solutions and semi-classical CFT
states starting from the intuition built by studying the \nBPS{4} solutions and working our way backwards. As described above, we use the
orbifold point language and our proposal for the \nBPS{8}  dictionary
is: 
\begin{quote}
\it A \nBPS{8} solution in supergravity describing a finite fluctuation
with modes $\hat{v}_{k,m}$ given in (\ref{modelattice}) around
AdS$_3\times S^3$ corresponds in the CFT to a semi-classical state
composed of strands of different types. The types of strands considered in this paper are 
characterized by the length, $k$, the (left-moving) momentum number, $m$,
and the choice of fermion ground state $( (0, 0)$ or $(\half, \half))$.  
The frequency with which each type of strand appears in the
CFT state must be large, and corresponds in the bulk to how much the
parameters of the supergravity solution (such as the Fourier
coefficients in $Z_I$ and $\Theta_I$) differ from those of AdS$_3\times
S^3$ (given in (\ref{supertube})).
\end{quote}
Clearly the novelty as compared to the two-charge states is the appearance
of a new quantum number $m$ determining the momentum of each type of
strand. In general, the momentum is carried by all possible types of
excitation that are available in the CFT and, in particular, on a strand
of length $k$, we can have modes of the free boson and fermion fields
carrying a fractional quantum of momentum in units of the inverse of the radius $R$. The usual orbifold rules only constrain the total momentum on each strand to be
integer-valued\footnote{Note however the important fact that this
orbifold-CFT rule, that each strand carries integral units of momentum,
must be refined in certain situations in the D1-D5 CFT; see
\cite[Section 6.3]{Giusto:2004ip} for more detail.  Here we ignore this
point and only consider integral units of momentum on each strand. }.

For general momentum-carrying states on a strand, it seems quite
non-trivial to determine the precise correspondence between the
frequency with which the strand appears in the CFT state and the deformation
parameter of the supergravity data.  However, for the particular ansatz
we consider here, the dictionary is simple enough and
can be inferred from the data we have collected.

As discussed in the gravity part of this paper,  we have focussed on a
class of states in which the momentum excitations are the operators
$(J_{-1}^+)_\ell^{}$ acting on different groups of $(0,0)$
strands (recall that the subscript $\ell$ means that this operator belongs to the $\ell^{\rm th}$ strand). This is the same type of strand that, in absence of momentum
carrying excitations, is related to the $g_5$ component of the
profile~\eqref{circleplus}. Thus it is natural to relate the presence of
this type of strand to the presence of a term $Z_4^{(k,m)}$ (see
 \eqref{Z4modes}) in the supergravity solution. As a consistency
check, if we set $m=0$, then this correctly reduces to the dictionary
discussed above for the \nBPS{4}  states: on the CFT side this kills the
momentum-carrying excitations and on the gravity side we recover the solution
\eqref{supertubeplus} which can be obtained from the profile from~\eqref{circleplus} by using
the general relations summarized in~\eqref{generaltwocharge}.

\begin{figure}[htbp]
\begin{quote}
   \begin{center}
    \vspace*{.5cm}
 \hspace*{-1cm}\includegraphics[height=2.8cm]{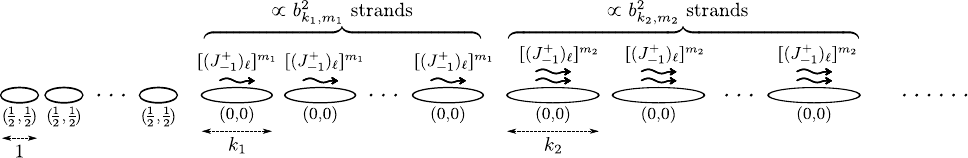}
 \caption{\label{fig:RR_gnd2} \sl 
 The dual CFT state for the superstrata geometries in Section
 \ref{Sect:examples}. The symbol  $[(J_{-1}^+)_\ell^{}]^m$ above a $(0,0)$ strand means that
 we act $m$ times by the operator $(J_{-1}^+)_\ell^{}$ on the ground state
 of the CFT on the strand.  The wavy arrows represent the non-vanishing
 momentum modes excited on the strand. The number of strands of the same
 type is $\cO(N)$, meaning that our solution represents a finite non-linear deformation
 around the AdS$_3\times S^3$ background. For the precise numbers of each type of strand, see \eqref{eq:Nkm}.}
   \end{center}
\end{quote}
\end{figure}

It is now straightforward to characterize the states that are dual to
the superstrata geometries we constructed: it is sufficient to look at
the form of $Z_4$ in~\eqref{Z4full} and interpret each term of the sum
as indicating the presence of ${\cal N}_{k,m}$ strands (on average) of the type
$(j_\ell,\jt_\ell)=(0,0)$ with $m$ units of momentum carried by
$[(J_{-1}^+)_\ell^{}]^m/m!$, with
\begin{equation}
  \label{eq:Nkm}
  {\cal N}_{k,m} = {\cal N} \binom{k}{m}^{\!\!-1} {(b_{k,m})^2\over 2k}~,~~~
  {\cal N}=\frac{R^2 N}{Q_1 Q_5}~.
\end{equation}
We will also identify the number of strands of the type $(j_\ell,\jt_\ell)=(\half,\half)$ with ${\cal N} a^2$. The numerical factors are suggested by the
supergravity expressions for the charges derived in
Section~\ref{Sect:physics}. First, in our superstratum state the average numbers of strands of winding $k$ multiplied by $k$ should sum up to the total number of CFT copies:
\begin{equation}
  \label{eq:totNcft}
  N = {\cal N} \left[a^2 + \sum_{(k,m)} k \binom{k}{m}^{\!\!-1} 
		\frac{(b_{k,m})^2}{2k}\right]~.
\end{equation}
This relation matches~\eqref{eq:q1q5ab}. Also the angular momentum charges~\eqref{eq:JtJgr} can be easily checked from the microscopic picture: only the first type of strands in Figure~\ref{fig:RR_gnd2} carries right-moving angular momentum.  Since the number of such strands is proportional to $a^2$, this matches the second relation in~\eqref{eq:JtJgr}.  The first relation in this equation follows from the fact that strands with $[(J_{-1}^+)_\ell^{}]^m$ in Figure~\ref{fig:RR_gnd2} carry $m$ units of left-moving angular momentum $j$, while the fermion zero modes of the strand $(\half,\half)$ contribute $\half$ each to $j$. Similarly, ~\eqref{eq:npgr} is consistent with the fact the operator $[(J_{-1}^+)_\ell^{}]^m$ adds $m$ units of momentum on each strand where it acts. Indeed, \eqref{eq:npjj} shows that one quantum of momentum is associated with one quantum of angular momentum and so each such quantum must be created by some $(J_{-1})_\ell$.

The identification between each single term $Z_4^{(k,m)}$ and the presence of many copies of an excited type of strand is also supported by some general properties of the momentum-carrying operator we used. For instance, by using the free orbifold description of the CFT it is possible to see that $[(J_{-1}^+)_\ell^{}]^m$ vanishes when it acts on a strand of length $k$ if $m>k$. This can be easily verified in the orbifold CFT\@.  One can simply note that this is a standard null vector identity in a current algebra of level $k$, or one can  use~\eqref{eq:psiell} and~\eqref{Jdefn} directly. The terms in $(J_{-1}^+)_\ell^{}$ that act non-trivially on the $(0,0)$ strands have the form $(\psi^{+\dot{1}}_{\frac{n}{k}})_\ell^{} (\psi^{+\dot{2}}_{1-\frac{n}{k}})_\ell^{}$ with $0\leq n \leq k$; so in $[(J_{-1}^+)_\ell^{}]^m$ with $m>k$ at least one fermionic creation operator appears twice, which implies that only strands with $m\leq k$ are possible. As one can see from equation~\eqref{eq:Deltakm} exactly the same constraint arises on the supergravity side as a regularity condition for $\Delta_{k,m}$. 

Another check that one can perform is to choose very particular values for the parameters defining the superstratum states so as to reconstruct a descendant state. For instance, in Figure~\ref{fig:RR_gnd2} we can consider just a single type of momentum-carrying strand with $m=k$. By generalizing the CFT analysis of~\cite{Giusto:2013bda}, one can check that this is a descendant obtained by choosing $\chi=\pi/2$ in the rotation of Section~\ref{ss:newmodes}. Then the (average) number of momentum-carrying strands has to be equal to the (average) number of $(0,0)$ strands in the seed two-charge geometry of Section~\ref{ss:seed}, as it can be seen by putting $m=k$ in~\eqref{eq:Nkm}.

The main message of this construction is that the linearized (in the
parameters $b_{k,m}$) expression for the scalar fields $Z_i$ is
sufficient to identify the dual state on the CFT side. On the bulk
side the supergravity equations allow one (at least for this class of
states) to find the explicit non-linear solutions. At this point it is
possible to further check the dictionary between CFT states and
geometries by comparing the expectation values for the protected
operators as it was done for the \nBPS{4}  solutions
in~\cite{Kanitscheider:2006zf,Kanitscheider:2007wq} and for 
their \nBPS{8} descendants in~\cite{Giusto:2013rxa}.

To conclude this section we want to underline  the significance of the
three-charge supergravity solutions that we have built.
The three-charge geometries with a precise CFT dual that have been known
prior to this paper \cite{Giusto:2004id, Ford:2006yb, Lunin:2012gp,
Giusto:2013bda} have been obtained by a solution-generating technique
\cite{Mathur:2003hj} that amounts to applying $\cR$-symmetry generators
such as $J^+_{-1}$ on \nBPS{4} states.  This procedure can only generate
an extremely restricted class of momentum-carrying states.  In technical
terms, one can only obtain the $\cR$-current descendants of chiral
primaries.\footnote{There is minor abuse of terminology here: since the
D1-D5 CFT is in the RR sector, what we really mean by a chiral primary
is the spectral flow of a chiral primary in the NS-NS sector. The same
caveat applies to the subsequent discussions.\label{ftnt:chprR}}
In contrast, our geometries correspond to descendants of
\emph{non-chiral primaries} and specifically, states generated by the
small current algebras, (\ref{Jdefn}), on different types of strand. Our
approach thus yields completely new, broad classes of solutions.

From the CFT perspective, the way we have achieved this can be described more precisely  as follows: 
Our three-charge states, such as the one described in Figure
\ref{fig:RR_gnd2}, are composed of multiple strands, in each of which we
have applied the modes $(J^+_{-1})_\ell^{}$ on a \nBPS{4} ground state
(= chiral primary).  Such a strand on its own can be thought of as
representing a descendant of a chiral primary.  However, when we have
two or more strands, the full state is a tensor product of
descendants of chiral primaries.  Now recall that, although the
tensor product of chiral primaries is again a chiral primary, the tensor
product of descendants of chiral primaries is in general a descendant of
a \emph{non-chiral primary} \cite{deBoer:1998ip}.  Therefore, a multi-strand
state in general represents a descendant of a non-chiral primary.  This
is because a tensor product of strands each of which is acted by
$(J^+_{-1})_\ell^{}$ cannot generally be written as the $\cR$-symmetry
generator $J^+_{-1}=\sum_\ell (J^+_{-1})_\ell^{}$ acting on a chiral
primary, except for special states in which numbers of different types
of strand are tuned in some precise way.

We can see the same physics from the supergravity perspective:  In our solution we allowed for arbitrary coefficients $b_{k,m}$
in the linear combination in~\eqref{Z4full}. This means that, in
general, it will not be possible to rewrite the solution as an element
of the $\cR$-symmetry group, which is an exponential of the operator $J^i =
\sum_\ell (J^i)_\ell^{}$, acting on a two-charge solution (= chiral
primary).
Descendant states such as the ``rigidly-generated''  solution discussed in
subsection \ref{ss:particularSoln} or those considered
in~\cite{Giusto:2013rxa, Giusto:2013bda} appear as special cases where
the coefficients of the various terms in the expression for the $Z_I$
are chosen in a precise way that allows one to reconstruct the currents $J^i$.

\section{Discussion, conclusions and outlook}
\label{Sect:Concl}

First and foremost we have constructed an example of a superstratum with
sufficient genericity to substantiate the claim that the superstratum
exists within supergravity as a smooth solution parameterized by at least one function of two
variables.  This, in itself, represents huge
progress within the programme of reproducing the black-hole entropy by counting microstate geometries that are valid in the same regime of parameters where the classical black hole solution exists, and is cause enough for the ``white smoke'' and celebration suggested by this
paper's title.  At a more technical level we have given an 
algorithm that can be effectively implemented to generate shape modes of
the superstratum.

We have also begun to develop a systematic picture of the holographic
duals of our superstrata and the results presented here contain several
important new results: Up until now, all the three-charge geometries constructed by solution generating methods starting from two-centered  geometries \cite{Giusto:2004id,Ford:2006yb, Lunin:2012gp, Giusto:2013bda} were dual to  descendants of  \emph{chiral primary states}\footnote{See footnote \ref{ftnt:chprR}.} in the left-moving sector of the D1-D5 CFT\@.  To obtain
the most general \nBPS{8} state one must be able to find the gravity
duals of arbitrary left-moving states: descendants of \emph{non-chiral
primaries}.  We have shown how such states are indeed being captured by
the superstratum.

Our focus in this paper has been to exhibit one example of a
superstratum rather than attempt an analysis of the possible families of
superstrata.  As a result we have passed over many interesting and
important physical and mathematical issues that arise from our
construction and we would like to catalog some of them.

\bigskip

We begin with the interpretation of our work in terms of the CFT\@. 
In  \cite{Bena:2014qxa}, three of the present authors conjectured that
the fluctuations of the superstratum that are visible in six-dimensional
supergravity correspond to current algebra excitations  of the CFT\@.  
The current algebra in question is
generated by the modes $(J^i_n)_\ell^{}$ of (\ref{Jdefn}) acting on \emph{individual} strands
labeled by $\ell$, and the associated sector of the CFT has central
charge $c=N$ which is large enough to reproduce the asymptotic growth of
the entropy of the three-charge black hole.  This is in stark contrast
to the $\cR$-symmetry algebra generated by the total
$J^i_n=\sum_\ell(J^i_n)_\ell^{}$ whose central charge is merely $c=\frac{3N}{N+2} < 3$. 
Therefore, in order to understand the fluctuation modes of the
superstratum and reproduce the black-hole entropy growth, it is crucial
to study how individual generators $(J^i_n)_\ell^{}$ are realized in
supergravity and whether they generate smooth geometries.

The solution generating technique that was used in the literature  \cite{Giusto:2004id,
Ford:2006yb, Lunin:2012gp, Giusto:2013bda}
to obtain smooth three-charge solutions starting from two-center geometries constructs solutions that are descendants of a chiral primary by the action of the total generator
$J^i_n$, and thus does not allow one to change independently each individual 
$(J^i_n)_\ell^{}$.  However, by taking a tensor product of such descendant
states, which corresponds in supergravity to taking a linear
superposition\footnote{This superposition can be understood as follows.  If one has a
free harmonic oscillator with the annihilation operator $a$, a classical
configuration with amplitude $\alpha$ corresponds to the coherent state
$e^{\alpha a^\dagger}\ket{0}\equiv \ket{\alpha}_a$.  If one has two
oscillators with $a$ and $b$, the classical configuration in which the
$a$ oscillator with amplitude $\alpha$ and the $b$ oscillator with
amplitude $\beta$ are classically superposed corresponds to the tensor
product state $\ket{\alpha}_a\otimes \ket{\beta}_b = e^{\alpha a^\dagger+\beta
b^\dagger}\ket{0}_{a,b}$.} and non-linearly completing it, we
successfully constructed smooth momentum-carrying geometries dual to
states that are not the result of acting on chiral primaries with the total $J^i_n$ but
intrinsically involve individual generators $(J^+_{-1})_\ell^{}$.
These explicit solutions demonstrate that the action of some individual
generators are indeed realized as smooth geometries.
We regard this as evidence in support of the conjecture that there exist
smooth supergravity solutions realizing the entire $c=N$ current algebra
generated by the individual generators $(J^i_n)_\ell^{}$.
Furthermore, the fact that our solutions involve two parameters $k,m$
suggests that the general fluctuation of the superstratum is described
by functions of at least two variables, as claimed in \cite{Bena:2014qxa}.

Although this represents major progress toward showing that the action of the full algebra of individual generators $(J^i_n)_\ell^{}$ gives smooth superstrata in gravity, there are still two more steps needed to achieve this goal: One must study how \emph{higher} generators $(J^i_{-n})_\ell^{}$ with $n\ge 2$ are realized in supergravity.  Furthermore, on a strand of
length $k$, we can have \emph{fractional} generators $(J^i_{-{n/
k}})_\ell^{}$ with $n\in \bbZ$; we must also study the bulk realizations of
these modes.
On a strand of length $k=1$, higher modes can account for the
three-charge black hole entropy growth $S\sim \sqrt{n_1 n_5 n_p}$ if
$n_p\gg n_1 n_5 $, while on a strand of length $k=N$, fractional modes can
account for the entropy growth if $n_1 n_5 n_p\gg 1$.  Therefore, either
higher modes or fractional modes are separately sufficient to reproduce
the three-charge black hole entropy for  large enough $n_p$.

We can look at these issues with the modes $(J^i_n)_\ell^{}$ from a
different angle.
As we have argued, the smooth geometries we have constructed represent
the tensor product of the descendants obtained by acting on chiral
primaries with the specific generator, $(J^+_{-1})_\ell^{}$ .  Actually,
the tower of descendant states built using $(J^+_{-1})_\ell^{}$ on a
chiral primary have a bulk interpretation as states of a supergraviton
\cite{Maldacena:1998bw}.  When there are multiple supergravitons, the
total state is the tensor product of such supergraviton states.
Therefore, the smooth geometries constructed in this paper must
correspond to the states of a supergraviton gas in the bulk.  More
precisely, our geometries can be regarded as coherent states in the
multi-particle Hilbert space of supergravitons.  Conversely, we expect
that the quantization of our solutions reproduce the multi-particle
supergraviton Hilbert space around AdS$_3\times S^3$.\footnote{The full
tower of descendants of a chiral primary involves the action of $L_{-1}$
and supercurrent generators, which we did not consider in this paper. To
reproduce the full Hilbert space, one needs to include the bulk
geometries generated by these generators too.}

\begin{figure}[htbp]
\begin{quote}
 \begin{center}
 \includegraphics[height=7cm]{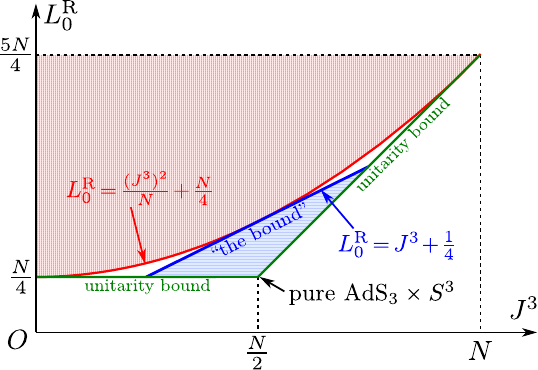}
 \caption{\label{fig:phase_diag} \sl 
The $J^3$-$L_0^{\rm R}$ phase diagram of the D1-D5 system. Pure
AdS$_3\times S^3$ corresponds to the point $(J^3,L_0^{\rm R})=({N\over
2},{N\over 4})$, and states exist only on and above the unitarity bound
(green lines).  The CFT elliptic genus can be reproduced by the bulk
graviton gas for $L_0^{\rm R}\le J^3+{1\over 4}$ (blue horizontally-hatched region).  Single-center BMPV black holes exist for $L_0^{\rm
R}\ge {(J^3)^2\over N}+{N\over 4}$ (red vertically-hatched region).  Even
in the region $L_0^{\rm R}< {(J^3)^2\over N}+{N\over 4}$, there exist
multi-center configurations of black holes and rings with a finite
horizon area \cite{Bena:2011zw}.
 }
 \end{center}
\end{quote}
\end{figure}

It was shown in \cite{deBoer:1998us, Maldacena:1999bp} that the
supergravity elliptic genus computed by counting these supergravitons,
with a stringy exclusion principle imposed by hand, agrees with the CFT
elliptic genus in the parameter region $L_0^{\rm NS}\le{N+1\over 4}$, or
in the R sector, $L_0^{\rm R}\le J^3+{1\over 4}$. (See 
Fig. \ref{fig:phase_diag}.)  In other words, in this parameter region, it
has been shown that every CFT state has a bulk realization as a
multi-supergraviton state, modulo the fact that some states are missed
because elliptic genus counts states with signs (it is only an index).  Therefore, our
geometries must be giving the bulk semi-classical description of all CFT
states in this parameter region (again, modulo possibly missed states).
However, this observation also illuminates what states our solutions
fail to capture.  The results in \cite{deBoer:1998us,
Maldacena:1999bp} imply that, above the bound $L_0^{\rm R}={J^3}+{1\over
4}$, the supergraviton gas is not enough to account for the CFT states.
By construction, the supergraviton gas includes neither higher
nor fractional modes and so we need these modes to reproduce the entropy
above the bound.  In particular, because single-center supersymmetric black
holes (the BMPV black hole \cite{Breckenridge:1996is}) exist above 
the bound,  we certainly need to understand superstratum
realizations of higher and/or fractional modes to reproduce entropy of this black
hole.

Actually, the story is even more interesting, since in 
\cite{Bena:2011zw} it was shown that there are multi-center black holes
(``moulting black holes'') even below the bound.  These black holes must
correspond to higher and/or fractional modes that are not visible in the
elliptic genus because of cancellations between bosonic and fermionic
states.  Therefore, understanding superstratum realizations of higher
and/or fractional modes are important also for understanding the
microstates of moulting black hole configurations.  The microstates of
the moulting black holes may be promising for studying higher and
fractional modes, because they exist even in the neighborhood of pure
AdS$_3\times S^3$ ($(J^3,L_0^{\rm R})=({N\over 2},{N\over 4})$).
Presumably, we can study those modes by looking at small deformations
around AdS$_3\times S^3$.

Returning to the supergravity perspective,  some important new ingredients 
are needed to make  further progress in the construction of the most
general supergravity superstrata.

As we noted above, the solutions constructed in this paper are based on the action of the
generators $(J_{-1}^-)_\ell^{}$. To go beyond this, we must  understand
supergravity realizations of $(J_{-n}^i)_\ell^{}$ with $n>1$ (higher modes)
and $n\in{\bbZ/ k}$ (fractional modes).  The total generators
$J_{-n}^i=\sum_\ell (J_{-n}^i)_\ell^{}$ are realized in the bulk by
dressing $J_0^i$ with a $v$-dependent exponential factor
$e^{-i\sqrt{2}\,nv/R}$ \cite{Mathur:2011gz}.  In general, the 
action of the total $J_{-n}^i$ makes all kinds of quantities
$v$-dependent, including the base metric $ds_4^2$ and the 1-form $\beta$
\cite{Giusto:2013bda}.  Therefore, the individual generators,
$(J_{-n}^i)_\ell^{}$, must also produce a complicated $v$-dependence in the
solution.  In particular, the new parameter, $n$, introduced by this
procedure is expected to generalize the phase $\hat{v}_{k,m}$ in
\eqref{modelattice} to $\hat{v}_{k,m,n}$ depending on three parameters
and lead to a much broader class of three-charge solutions.

Clearly, it is particularly important to understand
how the fractional modes $(J_{-n/k}^i)_\ell^{}$, which exist on strands of
length $k>1$, are encoded  in supergravity.  For $k \sim \cO(N)$, the fractional modes mean that
the bulk geometry must have an energy gap as low as $\,\sim
1/N$.\footnote{BPS excitations can only have a gap of order $\cO(1)$,
because supersymmetry means that the excitation energy is equal to the
momentum number which is quantized to integers.  Non-BPS excitations, on
the other hand, are not subject to such constraints and their energy gap
can be as small as $\cO(1/N)$.}  Furthermore, much of the three-charge
entropy  comes from excitations on strands with $k\sim \cO(N)$.
Although we have superstrata with shape modes that are intrinsically
two-dimensional, the actual modes studied here \emph{do not} have the
very low energy gap $\,\sim 1/N$.  In the gravity dual, excitations with
this energy gap are known to come from fluctuations of ``deep, scaling''
geometries in which the wavelength of the fluctuation is approximately
the scale of the horizon
\cite{Bena:2006kb,Bena:2007qc,deBoer:2008zn,deBoer:2009un}.  There are
thus two ways we might find such superstrata: One could consider a
single, large superstratum with a large dipole moment, $k$, and hence a
very large order, $\ZZ_k$, orbifold singularity and then allow
multi-valued functions with the fluctuation spectrum\footnote{A similar approach was used in \cite{Giusto:2012yz} to construct a
restricted class of microstates containing fractional modes.}.  (Of course,
multi-valued excitations are not allowed in supergravity and therefore
they must be excited multiple times so that their wavefunction is single-valued.)
While such a configuration is technically singular, its physical meaning is still
understandable. Alternatively, one could ``completely bubble'' such a
configuration to a $k$-centered configuration  and then the lowest
energy fluctuation will be some collective mode of all the bubbles in
this configuration in some deep scaling limit.  While the latter would have no orbifold
singularities, its lack of symmetry might make analytical computations
prohibitively hard.

More generally, there remains an important conceptual issue in
microstate geometries: we know how to obtain modes with the energy gaps
$\,\sim 1/N$ in both the D1-D5 CFT and in the deep, scaling holographic
dual geometries, but a detailed understanding of precisely how these
dual states are related remains unknown.  As indicated above, part of
the story must involve resolving orbifold singularities and multi-valued
functions but, on the gravity side, it must also involve deep scaling
geometries.  We would very much like to understand the emergence of such
scaling geometries from the detailed matching in the holographic
dictionary.  Understanding this issue is going to be an essential part
of seeing how the CFT entropy is encoded in the bulk geometry.

\bigskip

In the construction presented here we also encountered new types of singularities that are more difficult to remove than the singularities that appear in the standard construction of five-dimensional microstate geometries.  In the latter, the removal of singularities was related to the removal of closed time-like curves and this could be achieved by adjusting the choices of homogeneous solutions to the linear system of equations underlying the BPS solutions.  Here we have found that the choice of homogeneous solutions is insufficient for the task of singularity removal: one also has to interrelate the otherwise independent sets of fluctuations in order to obtain non-singular solutions.  We also noted that these interrelationships are very similar to those required for smooth horizons in black rings with fluctuating charge densities \cite{Bena:2014rea}.
  The physical origins and the resolutions of these potential singularities remains unclear and in this paper we simply exploited a mathematical algorithm to remove such singularities.  We would  like to understand the origins of such singularities, classify the ways in which one  can cancel them and see if there is, indeed, some physical link of this to black-ring horizon smoothness. 

In this paper we have also focused on superstrata that are asymptotic to AdS$_3 \times  {\rm S}^3$.  This choice was made for two reasons: simplicity and holography. The removal of singularities is simpler if the space is asymptotic to AdS$_3 \times  {\rm S}^3$ and  such asymptotics is all one needs for the study of the states that are holographically dual to our superstrata.  More generally, we would like to construct and classify superstrata that are asymptotic to $\mathbb{R}^{4,1} \times S^1$ or   $\mathbb{R}^{3,1} \times T^2$.  This means that constant terms need to be introduced into some of the metric functions. As noted in Section \ref{Sect:physics}, the constructions of such solutions should involve only straightforward technical issues rather than serious conceptual or physical issues.  Indeed, such solutions will be investigated in \cite{BGRSW:2015}.

There are also some other interesting technical issues in the mathematics of superstrata.  First, we found in Sections  \ref{Sect:layer2}   
and \ref{Sect:examples} that regularity required our Fourier coefficients to satisfy a quadratic constraint and that constraint came from canceling a class of  terms appearing in the quadratic $Z_1 Z_2 - Z_4^2$. Motivated by solution generating methods  \cite{Giusto:2004id,
Ford:2006yb, Lunin:2012gp, Giusto:2013bda} we chose to do this by leaving $Z_2$ and $\Theta_1$ unmodified (see (\ref{Z2full})) and adjusting the modes of $Z_1$ and $\Theta_2$ to cancel the problematic terms arising out of $Z_4^2$.  If one allows for a general set of modes in $Z_2$ and $\Theta_1$ then there are presumably many more ways to satisfy the quadratic constraints and hence more allowed excitations of the superstratum.  As we also noted, one must furthermore revisit the quadratic constraint on Fourier coefficients if one is to construct superstrata in asymptotically flat geometries and so we intend to analyze this constraint more completely in \cite{BGRSW:2015}.  

The second technical issue has to do with the existence of a systematic approach to solving the system of differential equations underlying our solutions. We have been in this paper able to completely solve for all the fields of the solution in closed form except for one function appearing in some of the components of the angular momentum vector\footnote{We have a solution for $\omega_\psi+ \omega_\phi$ and if we could find the function we miss we could find $\omega_\psi$ and $\omega_\phi$ independently, and then solve for $\omega_r$ and $\omega_\theta$ algebraically.}.  The two-centered system leads to some relatively simple differential operators and, in particular, the Laplacian (\ref{LapOp2}) is separable. The sources are also relatively simple functions and we have managed to find complete analytic solutions for some infinite families of sources. Our explicit solutions are also polynomials in simple functions of $r$ and $\theta$.   All of this suggests that there must be a far more systematic approach to solving this system of differential equations.  Indeed, we strongly suspect that the whole mathematical problem we have been solving in Section \ref{Sect:layer2} should have a much simpler formulation and solution in terms of some cleverly chosen orthogonal polynomials.  Understanding this may, in turn, lead to a clearer understanding of the whole system of differential equations and maybe even a reformulation of the general solution, perhaps even for multi-centered solutions, in terms of Green functions. Furthermore, solving this problem should enable the complete analytic construction of the most general superstratum based on two centers. Work on this is also continuing.

\section*{Acknowledgments}

We would like to thank Jan de Boer and Samir Mathur  for discussions.
The work of IB was supported in part by the ERC Starting Grant 240210
{\em String-QCD-BH}, by the National Science Foundation Grant
No.~PHYS-1066293 (via the hospitality of the Aspen Center for Physics)
by the John Templeton Foundation Grant 48222 and by a grant from the
Foundational Questions Institute (FQXi) Fund, a donor advised fund of
the Silicon Valley Community Foundation on the basis of proposal
FQXi-RFP3-1321 (this grant was administered by Theiss Research).
The work of SG was supported in part by the Padua University Project
CPDA144437. The work of RR was partially supported by the Science and
Technology Facilities Council Consolidated Grant ST/L000415/1 {\em
``String theory, gauge theory \& duality''}.
The work of MS was supported in part by Grant-in-Aid for Young
Scientists (B) 24740159 from the Japan Society for the Promotion of
Science (JSPS).
The work of NPW was supported in part by the DOE grant
DE-SC0011687.
SG, RR and NPW would like to thank Yukawa Institute for Theoretical
Physics for hospitality at the ``Exotic Structures of Spacetime''
workshop (YITP-T-13-07) during the early stages of this project.
SG, RR, MS and NPW are very grateful to the IPhT, CEA-Saclay for
hospitality while a substantial part of this work was done.
MS would like to thank the high energy theory group of the University of
Padua where this work was completed for hospitality.

\appendix

\section{D1-D5 geometries}
\label{Appendix:2charge}

The \nBPS{4} D1-D5 geometries invariant under $T^4$ rotations are associated with a profile $g_A(v')$ with non-trivial components for $A=i=1,\ldots,4$ and for $A=5$. Given such a profile, the functions and fields describing the geometry in the language of the IIB solution (\ref{ansatzSummary}) are
\begin{subequations}\label{generaltwocharge}
\begin{align}
& Z_2 = 1 + \frac{Q_5}{L} \int_0^{L} \frac{1}{|x_i -g_i(v')|^2}\, dv'~,~~~
  Z_4 = - \frac{Q_5}{L} \int_0^{L} \frac{\dot{g}_5(v')}{|x_i -g_i(v')|^2} \, dv' \,,\\\label{Z1profile}
& Z_1 = 1 + \frac{Q_5}{L} \int_0^{L} \frac{|\dot{g}_i(v')|^2+|\dot{g}_5(v')|^2}{|x_i -g_i(v')|^2} \, dv' ~, ~~~ d\gamma_2 = *_4 d Z_2~,~~~d\delta_2 = *_4 d Z_4~,\\
& A = - \frac{Q_5}{L} \int_0^{L} \frac{\dot{g}_j(v')\,dx^j}{|x_i -g_i(v')|^2} \, dv' ~, ~~~ dB = - *_4 dA~,~~~ ds^2_4 = dx^i dx^i~, \\
& \beta = \frac{-A+B}{\sqrt{2}}~,~~~\omega = \frac{-A-B}{\sqrt{2}}~,~~~{\cal F}=0~,~~~a_1=a_4=x_3=0~,
\end{align}
\end{subequations}
where the dot on the profile functions indicates a derivative with respect to $v'$ and $*_4$ is the dual with respect to the flat $\mathbb{R}^4$ metric $ds^2_4=dx^i dx^i$.
The D1 charge is given by
\begin{equation}
 Q_1={Q_5\over L}\int_0^L \bigl(|\dot{g}_i(v')|^2+|\dot{g}_5(v')|^2\bigr)dv'.
\end{equation}
The quantities $Q_1$, $Q_5$ are related to quantized D1 and D5
numbers $n_1$, $n_5$ by
\begin{equation}
Q_1 = \frac{(2\pi)^4\,n_1\,g_s\,\alpha'^3}{V_4}\,,\quad Q_5 = n_5\,g_s\,\alpha'\,.
\label{Q1Q5_n1n5}
\end{equation}
where $V_4$ is the coordinate volume of $T^4$.

\section{Solution of the (generalized) Poisson equation}
\label{Appendix:recursion}

The function $F_{k,m}^{(p,q)}$ was defined in the main text to be the
regular solution to equation \eqref{eq:poisson}, which we repeat here for
convenience:
\begin{equation}
 \widehat{\mathcal{L}}^{(p,q)} F_{k,m}^{(p,q)} = \frac{1}{r^2+a^2}\frac{\Delta_{k,m}}{\cos^2\theta\,\Sigma}\,,
  \label{eqFkmpqApp}
\end{equation}
where the generalized Laplacian $\widehat{\mathcal{L}}^{(p,q)}$ was
defined in \eqref{LapOp1}.  In this Appendix, we derive the explicit
solution to this equation, given in the main text in equation \eqref{explFkmpq}.

First, let us define the functions
\begin{align}
 G_{km}&\equiv {1\over r^2+a^2}\Delta_{k,m},\qquad
 S_{km}\equiv {1\over r^2+a^2}{\Delta_{k,m}\over \cos^2 \theta\,\Sigma}.
\label{defGkmSkm}
\end{align}
It is easy to show that these satisfy the following recursion
relation:
\begin{align}
 \cLh^{(p,q)} G_{km}=
 [p^2-(k+2)^2]S_{k+2,m+2}+[(k-m)^2-(p-q)^2]S_{k,m+2}
 +(m^2-q^2)S_{k,m}.
 \label{GSrecursion}
\end{align}
Now, let us introduce the following generating functions:
\begin{align}
\begin{split}
  \cF(\kappa,\mu)\equiv \sum_{k,m} F_{k,m}^{(p,q)}e^{k\kappa+m\mu},\quad
  \cG(\kappa,\mu)\equiv \sum_{k,m} G_{k,m}e^{k\kappa+m\mu},\quad
  \cS(\kappa,\mu)\equiv \sum_{k,m} S_{k,m}e^{k\kappa+m\mu}.
\end{split}
\end{align}
In terms of these, the equation we want to solve, \eqref{eqFkmpqApp},
can be collectively written as
\begin{align}
 \cLh^{(p,q)}\cF(\kappa,\mu)=\cS(\kappa,\mu),\label{cF,cS}
\end{align}
and the  recursion relation
\eqref{GSrecursion}  as
\begin{align}
 \cLh^{(p,q)}\cG(\kappa,\mu)
 &=
  \Bigl[-e^{-2\kappa-2\mu}(p^2-\partial_\kappa^2)
 +e^{-2\mu}((\partial_\kappa - \partial_\mu +2)^2-(p-q)^2)
 +(\partial_\mu ^2-q^2)\Bigr]\cS(\kappa,\mu).\label{cG,cS}
\end{align}
Because $\cLh^{(p,q)}$ commutes with $\partial_\kappa$ and
$\partial_\mu$, a comparison between \eqref{cF,cS} and \eqref{cG,cS}
gives
\begin{align}
 \cF(\kappa,\mu)
 &=
  \Bigl[-e^{-2\kappa-2\mu}(p^2-\partial_\kappa^2)
 +e^{-2\mu}((\partial_\kappa - \partial_\mu +2)^2-(p-q)^2)
 +(\partial_\mu ^2-q^2)\Bigr]^{-1}\cG(\kappa,\mu)
 \notag \\
 &=
 -\left[
 1-e^{2\kappa}{(\partial_\kappa-\partial_\mu+2)^2-(p-q)^2\over (\partial_\kappa+2)^2-p^2}
 -e^{2\kappa+2\mu }{\partial_\mu^2-q^2\over (\partial_\kappa+2)^2-p^2}
 \right]^{-1}
 \notag\\
 &\hspace{45ex}
 \times
 e^{2\kappa +2\mu }{1\over (\partial_k+2)^2-p^2}\cG(\kappa,\mu)
 \notag \\
 &=
 -\sum_{n=0}^\infty \left[
 e^{2\kappa}{(\partial_\kappa-\partial_\mu+2)^2-(p-q)^2\over (\partial_\kappa+2)^2-p^2}
 +e^{2\kappa+2\mu }{\partial_\mu^2-q^2\over (\partial_\kappa+2)^2-p^2}
\right]^n
 \notag\\
 &\hspace{45ex}
 \times
 e^{2\kappa +2\mu }{1\over (\partial_k+2)^2-p^2}\cG(\kappa,\mu).
\end{align}
By expanding the $n$th power using binomial coefficients and explicitly
writing down the first few terms in terms of $F_{k,m}^{(p,q)}$ and
$G_{k,m}$, one finds
\begin{align}
 F_{k,m}^{(p,q)}
 &=-\sum_{s=0}^\infty \sum_{t=0}^s
 {s\choose t}
 {
 \,\overbrace{(k-m+p-q)(k-m+p-q-2)\cdots}^{s-t}\,
 \overbrace{(m+q-2)(m+q-4)\cdots}^{t}\,
 \over
 \underbrace{(k+p)(k+p-2)\cdots}_{s+1}
 }
 \notag\\[-1.5ex]
 &\qquad
 \times
 {
 \,\overbrace{(k-m-p+q)(k-m-p+q-2)\cdots}^{s-t}\,
 \overbrace{(m-q-2)(m-q-4)\cdots}^{t}\,
 \over
 \underbrace{(k-p)(k-p-2)\cdots}_{s+1}
 }
 G_{k-2s-2,m-2t-2}
 \notag\\
 &=-{1\over k^2-p^2}
 \sum_{s=0}^{\infty}\sum_{t=0}^s
 {s\choose t}
 \frac
 {{{k+p\over 2} -s-1\choose {m+q\over 2} -t-1} {{k-p\over 2} -s-1\choose {m-q\over 2} -t-1}}
 {{{k+p\over 2} -1  \choose {m+q\over 2} -1  } {{k-p\over 2} -1  \choose {m-q\over 2} -1}}
 G_{k-2s-2,m-2t-2}.
 \notag\\
 &=-{1\over 4k_1 k_2(r^2+a^2)}
 \sum_{s=0}^{\infty}\sum_{t=0}^s
 {s\choose t}
 \frac
 {{k_1-s-1\choose m_1 -t-1} {k_2 -s-1\choose m_2 -t-1}}
 {{k_1  -1\choose m_1 -1  } {k_2   -1\choose m_2 -1}}
 \Delta_{k-2s-2,m-2t-2},
\end{align}
where the relations between $(k,m,p,q)$ and $(k_1,m_1,k_2,m_2)$ are given
in (\ref{kpm},\ref{system2}).  If we assume
\begin{align}
 k_1\ge m_1\ge 1,\qquad
 k_2\ge m_2\ge 1,
\end{align}
then the sum truncates at finite $s$ and $F_{k,m}^{(p,q)}$ can be written as
\begin{align} 
 F_{k,m}^{(p,q)} 
 &=-{1\over 4k_1 k_2(r^2+a^2)}
 \sum_{s=0}^{\min\{k_1,k_2\}-1}\sum_{t=0}^s
 {s\choose t}
 \frac
 {{k_1-s-1\choose m_1 -t-1} {k_2 -s-1\choose m_2 -t-1}}
 {{k_1  -1\choose m_1 -1  } {k_2   -1\choose m_2 -1}}
 \Delta_{k-2s-2,m-2t-2},
\end{align}
which is \eqref{explFkmpq} in the main text.  It is clear that  this
is a regular function at $r=0$.


\end{document}